\documentclass[twocolumn]{aastex63}
\usepackage{color}
\usepackage{CJKutf8}
\graphicspath{{./}}

\accepted{September 29, 2020}

\shorttitle{Radio emission during microflare}
\shortauthors{Sharma et al.}

\begin{document}

\title{Radio and X-ray Observations of Short-lived Episodes of Electron Acceleration in a Solar Microflare}

\newcommand{\cntext}[1]{\begin{CJK}{UTF8}{gbsn}#1\end{CJK}\kern-1ex}

\correspondingauthor{Rohit Sharma}
\email{rohit.sharma@fhnw.ch, rohitcbscient@gmail.com}

\author[0000-0003-0485-7098]{Rohit Sharma}
\affil{Fachhochschule Nordwestschweiz \\
Bahnhofstrasse 6, \\
5210 Windisch, Switzerland}

\author[0000-0003-1438-9099]{Marina Battaglia}
\affiliation{Fachhochschule Nordwestschweiz \\
Bahnhofstrasse 6, \\
5210 Windisch, Switzerland}

\author{Yingjie Luo}
\affiliation{Center for Solar-Terrestrial Research, New Jersey Institute of Technology, 323 M L King Jr Blvd, Newark, NJ 07102-1982, USA}

\author[0000-0002-0660-3350]{Bin Chen (\cntext{陈彬})}
\affiliation{Center for Solar-Terrestrial Research, New Jersey Institute of Technology, 323 M L King Jr Blvd, Newark, NJ 07102-1982, USA}

\author[0000-0003-2872-2614]{Sijie Yu (\cntext{余思捷})}
\affiliation{Center for Solar-Terrestrial Research, New Jersey Institute of Technology, 323 M L King Jr Blvd, Newark, NJ 07102-1982, USA}



\begin{abstract}
Solar flares are sudden energy release events in the solar corona, resulting from magnetic reconnection, that accelerates particles and heats the ambient plasma. During a flare, there are often multiple, temporally and spatially separated individual energy release episodes that can be difficult to resolve depending on the observing instrument.  
We present multi-wavelength imaging and spectroscopy observations of multiple electron acceleration episodes during a GOES B1.7-class two-ribbon flare on 2012 February 25, observed simultaneously with the Karl G. Jansky Very Large Array (VLA) at 1--2 GHz, the Reuven Ramatay High Energy Solar Spectroscopic Imager (RHESSI) in X-rays, and the Solar Dynamics Observatory in extreme ultraviolet (EUV). During the initial phase of the flare, five radio bursts were observed. A nonthermal X-ray source was seen co-temporal, but not co-spatial, with the first three radio bursts.  Their radio spectra are interpreted as optically thick gyrosynchrotron emission. 
By fitting the radio spectra with a gyrosynchrotron model, we derive the magnetic field strength and nonthermal electron spectral parameters in each acceleration episode. Notably, the nonthermal parameters derived from X-rays differ considerably from the nonthermal parameters inferred from the radio. The observations are indicative of multiple, co-temporal acceleration episodes during the impulsive phase of a solar microflare. The X-ray and radio burst sources likely originate from separate electron distributions in different magnetic loops.
\end{abstract}



\section{Introduction}
In a solar flare, energy stored in the coronal magnetic field is rapidly released. As a consequence, particles are accelerated to energies from a few tens of keV to MeVs and the ambient plasma is heated. X-ray bremsstrahlung emission and radio gyrosynchrotron emission provide diagnostics of the accelerated electrons and the ambient plasma conditions. 

Observations at both wavelength ranges are highly complementary, providing information on the spectral properties of flare-accelerated electrons in a broad energy range, as well as timing and location of the energetic electrons \citep[see][for a review]{2011SSRv..159..225W}. While the intensity of X-ray bremsstrahlung emission predominatly depends on the density of the ambient plasma and the nonthermal electron density, the intensity and spectral shape of gyrosynchrotron emission strongly depends on the magnetic field \citep[e.g.][]{1982ApJ...259..350D,2013SoPh..288..549G,Fleishman2020}. In the general solar flare scenario, electrons are accelerated in the corona at or above the top of a magnetic loop \citep{1994Natur.371..495M,2010ApJ...714.1108K}, one generally expects X-ray emission at nonthermal energies to originate from the dense chromosphere at the footpoints of the loop, while gyrosynchrotron emission at a few GHz presumably originates from electrons of the same population, most likely trapped in the loop \citep[e.g.][]{2011SSRv..159..225W}. The number of trapped nonthermal particles depends primarily on the mirror ratio, pitch angle, and particle diffusion into the loss cone \citep{2002ASSL..279.....B,2014ApJ...787..123H}.

Observations at both wavelength ranges frequently hint the presence of multiple acceleration episodes at different times and locations during a single flare, often originating from multiple reconnection sites. 
In microwaves, a variety of bursts and features that are interpreted as multiple acceleration episodes are frequently observed in dynamic spectra and light curves at time scales of milliseconds to seconds \citep[e.g.][]{1990SoPh..130..183A,1994A&AS..104..145I,1994ApJS...90..693S,2005HvaOB..29..309M,2010AdSpR..46..413Y,2015SoPh..290..287Z,2008AdSpR..41..936M,2012SoPh..280..537M,2016A&A...593A..80M}.  
Indications of multiple acceleration episodes during a flare have also been found in X-ray light curves as far back as \citet{1969ApJ...155L.117P}, and in recent times by e.g. \citet{2005ApJ...625L.143G} and \citet{2016ApJ...833..284I}, although with time scales of individual episodes usually ranging from a few seconds to a few tens of seconds. Occasionally, subsecond X-ray pulses have also been observed \citep{1983ApJ...273..783K,1984ApJ...287L.105K,1995ApJ...447..923A,2018ApJ...867...84G,2019ApJ...883...38A,2020arXiv200305007K}.

Spatially, the location of microwave sources associated with individual acceleration episodes can fluctuate rapidly. \cite{Kai1986} found displacements between 8$''$--12$''$ for different individual bursts within a few seconds. Similar changes in position have also been observed in X-rays \citep[e.g.][]{2016SoPh..291.3385K}. Where microwave and X-ray observations of the same event exist, the observed microwave bursts are often temporally associated with X-ray brightenings \citep{1981ApJ...251..342K,2015ApJ...807..124K}. However, despite the temporal association, X-ray and microwave sources may not be co-spatial \citep[e.g.][]{2015ApJ...807..124K,2018ApJ...852...32K,2018ApJ...867...84G}. 

One of the many advantages of modern microwave observations is their high temporal resolution. 
While many previous observations provided source locations at discrete frequencies, it has only recently become possible to investigate the detailed spectral evolution of flares as a function of location and time. 
In decimeter--centimeter wavelengths, such a capability, known as ``dynamic imaging spectroscopy'', is now provided by the Karl G. Jansky Very Large Array (VLA; \citealt{2011ApJ...739L...1P}) and the Expanded Owens Valley Solar Array (EOVSA; \citealt{2018ApJ...863...83G}). This allows us to investigate individual acceleration episodes spatially and spectrally at subsecond timescales  \citep[e.g.,][]{2013ApJ...763L..21C,2015Sci...350.1238C,2018ApJ...866...62C,2019ApJ...884...63C,2017ApJ...848...77W,2019ApJ...872...71Y}, a capability not currently available in X-rays. The Reuven Ramaty High Energy Solar Spectroscopic Imager \citep[RHESSI;][]{Li02} provided imaging spectroscopy capability between 2002 and 2018 at a nominal temporal resolution of four seconds, but in practice, imaging spectroscopy typically needs large integration times of the order of tens of seconds to minutes. Nevertheless, RHESSI provides crucial complementary diagnostic of the flaring processes observed with the VLA.  

Here we present comprehensive observations of multiple individual bursts during a GOES B1.7-class (background subtracted) microflare observed jointly in radio by the VLA, in X-rays by RHESSI, and in the Extreme Ultraviolet (EUV) by the Atmospheric Imaging Assembly \citep[AIA;][]{2012SoPh..275...17L} on board the Solar Dynamics Observatory \citep[SDO;][]{2012SoPh..275....3P}.  An event overview is given in Section~\ref{s:overview}. The radio and X-ray analysis, including imaging and spectral spectral fitting, is described in Section~\ref{s:analysis}. In Section~\ref{s:evol}, we describe the spatial evolution of the radio and X-ray sources. The findings are discussed in Section~\ref{s:discuss}, followed by conclusions in Section~\ref{s:conclusion}.

\section{Event overview}\label{s:overview}
The flare SOL2012-02-25T20:50:34 occurred in active region 11421.
\begin{figure*}
\begin{center}
\includegraphics[trim = 0cm 0cm 0.5cm 1.8cm,width=0.8\textwidth]{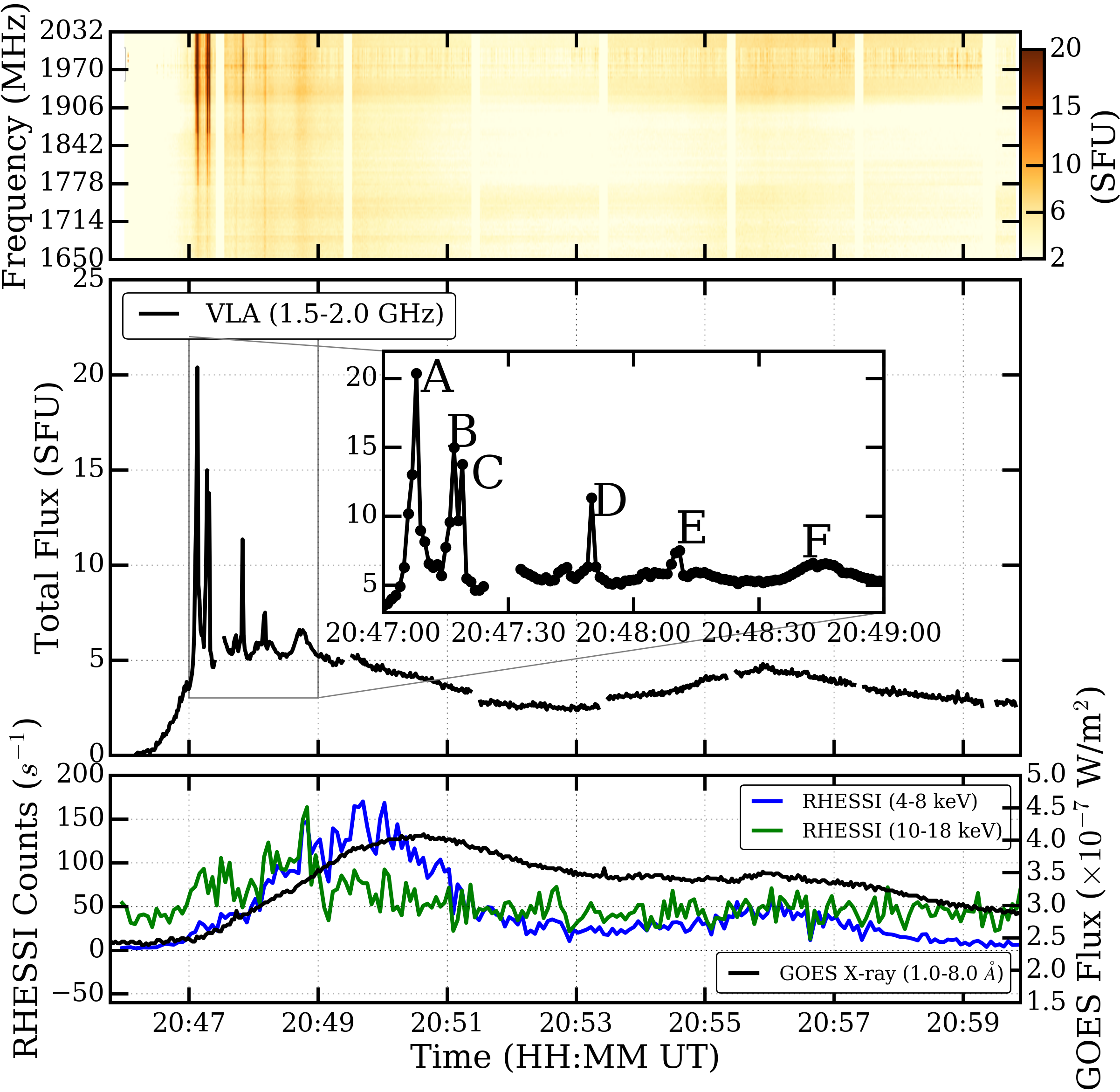}
\caption{Temporal evolution of radio and X-ray emission during the B1.7-class flare on 2012 February 25 (SOL2012-02-25T20:50:34). Top: VLA dynamic spectrum showing the total flux computed from the radio images for each frequency-time pixel in the observation. Each pixel has a size of 4 MHz and 1 s in frequency and time respectively.  The second panel shows the frequency averaged VLA spectrum from 1.65 GHz to 2.03 GHz. The inset shows 5 distinct radio bursts marked by letters. The third panel shows X-ray light curves from RHESSI and GOES.  \label{Fig:overview}}
\end{center}
\end{figure*}
Figure~\ref{Fig:overview} shows the radio dynamic total power spectrum observed by the VLA's L-Band receiver between 1.65 and 2.03 GHz, as well as X-ray light curves from GOES at 1--8 \AA\ and RHESSI at 4--8 keV and 10--18 keV. 
The radio emission begins to rise from 20:46 UT, followed by many short-lived bursts observed between 20:47--20:49 UT in 1.65 to 2.03 GHz frequency range. We identify six distinct radio bursts (labeled as A, B, C, D, E and F in Figure~\ref{Fig:overview}). The peak flux densities of individual bursts range from 7 to 20 SFU and decay over time. 
After 20:49:00 UT, enhanced continuum emission is seen until the end of the observation. The radio bursts coincide with impulsive emission at X-ray energies above 10 keV. 

GOES 1--8 \AA\ and RHESSI 4--8 keV X-ray light curves show one main peak at around 20:50 UT and a secondary brightening at around 20:56 UT. The higher-energy (10--18 keV) RHESSI X-ray light curve displays short-duration variations between 20:47--20:49 UT. The impulsive X-ray bursts are nearly co-temporal with the radio bursts in the time interval of interest (20:47--20:49 UT).

\begin{figure*}
    \centering
  \begin{tabular}{cc}
\includegraphics[trim = 0cm 0cm 0.0cm 0.0cm,width=0.38\textwidth]{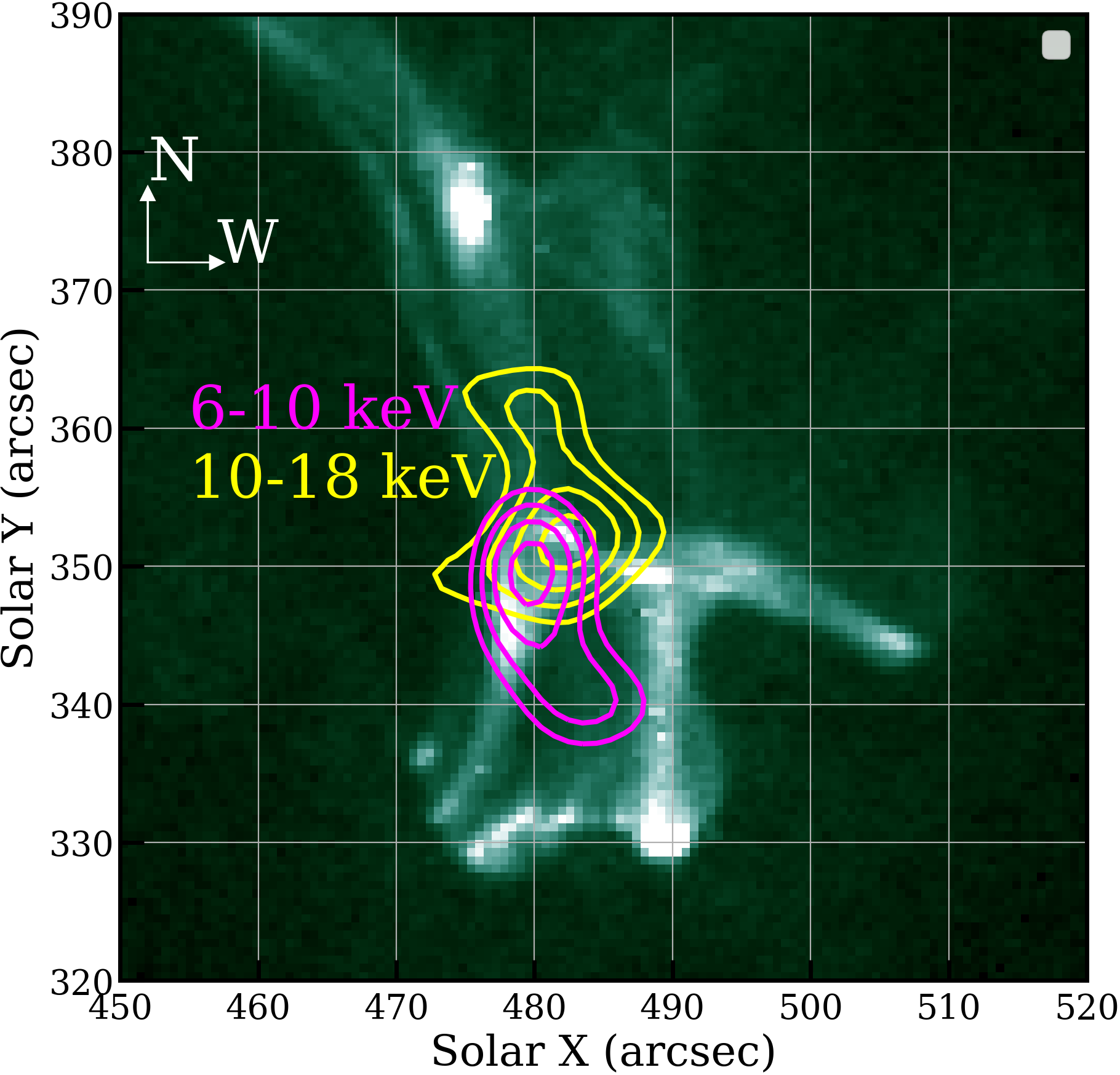}
&
\includegraphics[trim = 0cm 0cm 0.0cm 0.0cm,width=0.45\textwidth]{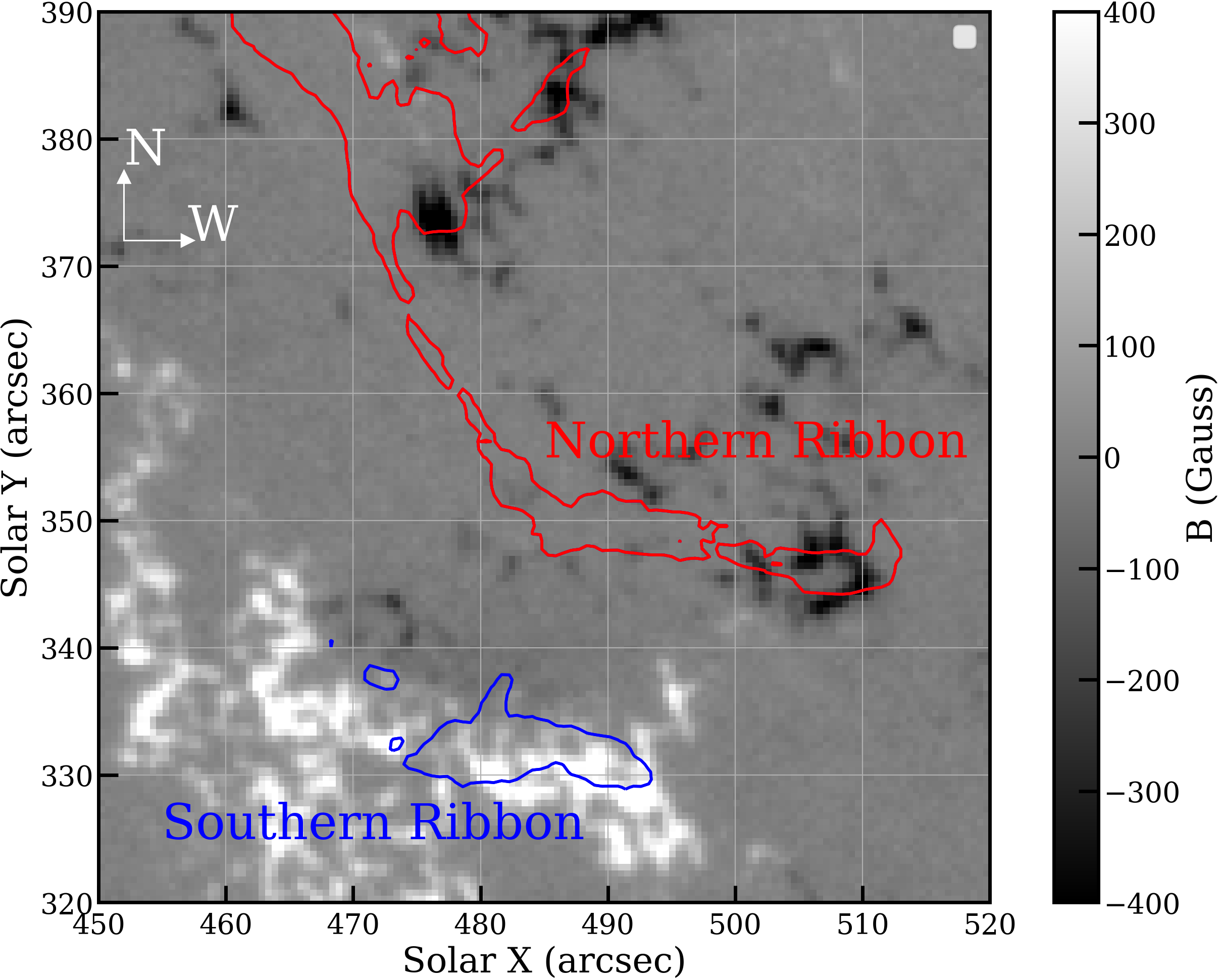}
\end{tabular}
    \caption{Overview of flare morphology and magnetic topology. Left panel: AIA 94 \AA \  image at 20:49:02 UT overlaid with RHESSI X-ray sources (magenta and yellow contours) integrated from 20:47:00 to 20:47:28 UT. The RHESSI contour levels are at 65\%, 75\%, 85\% and 95\% w.r.t the map's peak. Right panel:  HMI magnetogram at 20:51:45 UT. The red and blue contours are emission observed by AIA 304 \AA \ at 20:51:32 UT for northern and southern ribbons respectively. The contour level shown is at 18\% of the maximum brightness.}
    \label{Fig:ribbons}
\end{figure*}

Even though the event is weak, it is morphologically complex with two main ribbons connected by several loop systems, as shown in SDO/AIA and SDO/Helioseismic and Magnetic Imager (HMI) images (Figure \ref{Fig:ribbons}). The HMI image shows two regions of opposite magnetic polarity, which are co-spatial with two flare ribbons as seen in the AIA 304 \AA\ wavelength channel. The AIA 94 \AA\ image shows an arcade of loops that connects the western part of the two ribbons and an additional longer loop that connects the southern ribbon with the eastern end of the northern ribbon.

\section{Radio and X-ray analysis}\label{s:analysis}
The VLA observations had a frequency coverage of 1 GHz to 2.03 GHz ($\lambda =$ 15--30 cm) with a spectral resolution of 1 MHz and temporal resolution of 1 second in both the right-hand- and left-hand-circular polarization (RCP and LCP). The spectral range of the observation was divided uniformly into 8 spectral windows. Each spectral window had 128 1-MHz-wide frequency channels. The observations were taken in the C configuration of the VLA, which had a maximum baseline length of $\sim$3 km. A total of 27 antennas (i.e., the full array) were used for the observation. This provides an angular resolution, represented as the full width half maximum (FWHM) of the synthesized beam, of $15''$ by $10''$ at 2 GHz, which is inversely proportional to the observing frequency ($1/\nu_{\rm GHz}$). 
RHESSI observed this event in its standard observing mode. It had completed an annealing procedure of its germanium detectors three days prior to the present observations, which resulted in seven out of nine detectors being in optimal working condition with good sensitivity and spectral resolution. 


\subsection{X-ray imaging and spectral analysis}
\label{Sec:X-ray}
Using the standard RHESSI data analysis package in the IDL {\tt SolarSoftware}, we produced images and spectra between 20:46 UT and 20:59 UT, using 60 seconds time integration to ensure high enough count rates, except for the time interval between 20:47 UT and 20:48 UT which was split into two intervals lasting from 20:47:00 UT to 20:47:28 UT and from 20:47:28 UT to 20:48:00 UT. 
For X-ray imaging, data from detectors 3, 5, 6, 8 were used for image reconstruction with the {\tt CLEAN} algorithm \citep{Hur02}. For each time interval, images averaged over two energy-bands were made: 4--8~keV and 10--18~keV. A clean-beam-width factor of 1.4 was applied, resulting in a FWHM of the CLEAN beam of 9.6 arcsec.

The X-ray count spectrum of RHESSI detector 1 was used for spectral fitting with the standard {\tt OSPEX}  package \citep{2002SoPh..210...33S}. The fitting model consisted of a thermal component at low energies.  Between 20:47:00~UT and 20:50:00~UT, significant emission at higher energies was present that was fitted with an additional nonthermal thick-target component. The same time-integration was used as for imaging. 
\begin{figure}
\begin{center}
\includegraphics[trim = 0cm 0cm 0.0cm 0.0cm,width=0.45\textwidth]{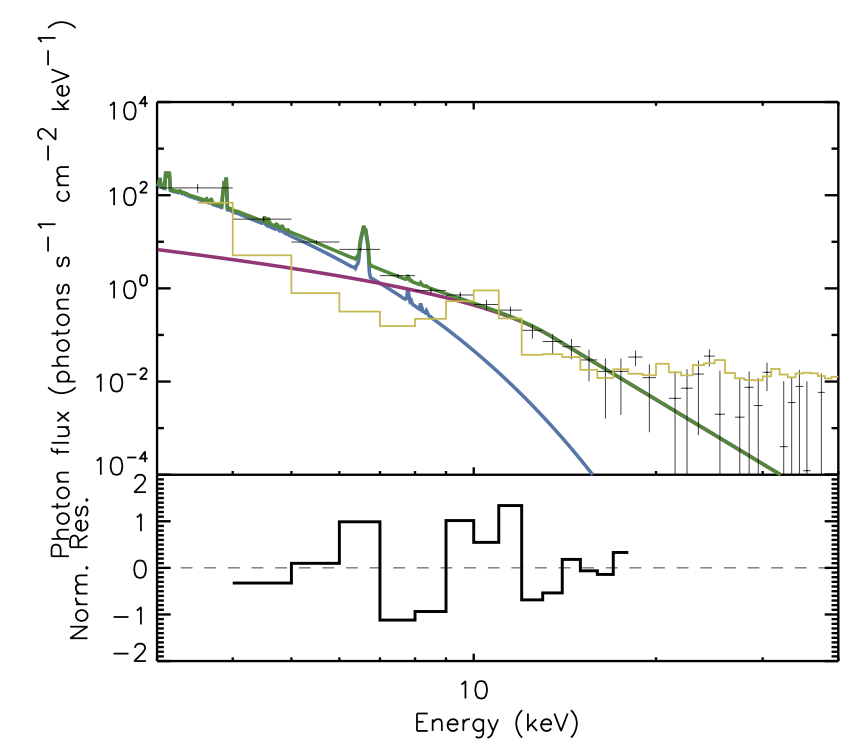}  
\caption{Top panel: RHESSI spectrum along with the fitted thermal and nonthermal components between 20:47:00 UT and 20:47:28 UT. The black crosses show the RHESSI spectrum, while the green curve shows the sum of the thermal (blue line) and nonthermal (purple line) components. The yellow histogram shows the background spectrum. Bottom panel: Normalised residuals between the spectrum and its fit. 
} \label{Fig:rhessi_spec}
\end{center}
\end{figure}

\begin{figure*}
    \centering
  \begin{tabular}{cc}
\includegraphics[trim = 0cm 0cm 0.0cm 0.0cm,width=0.5\textwidth]{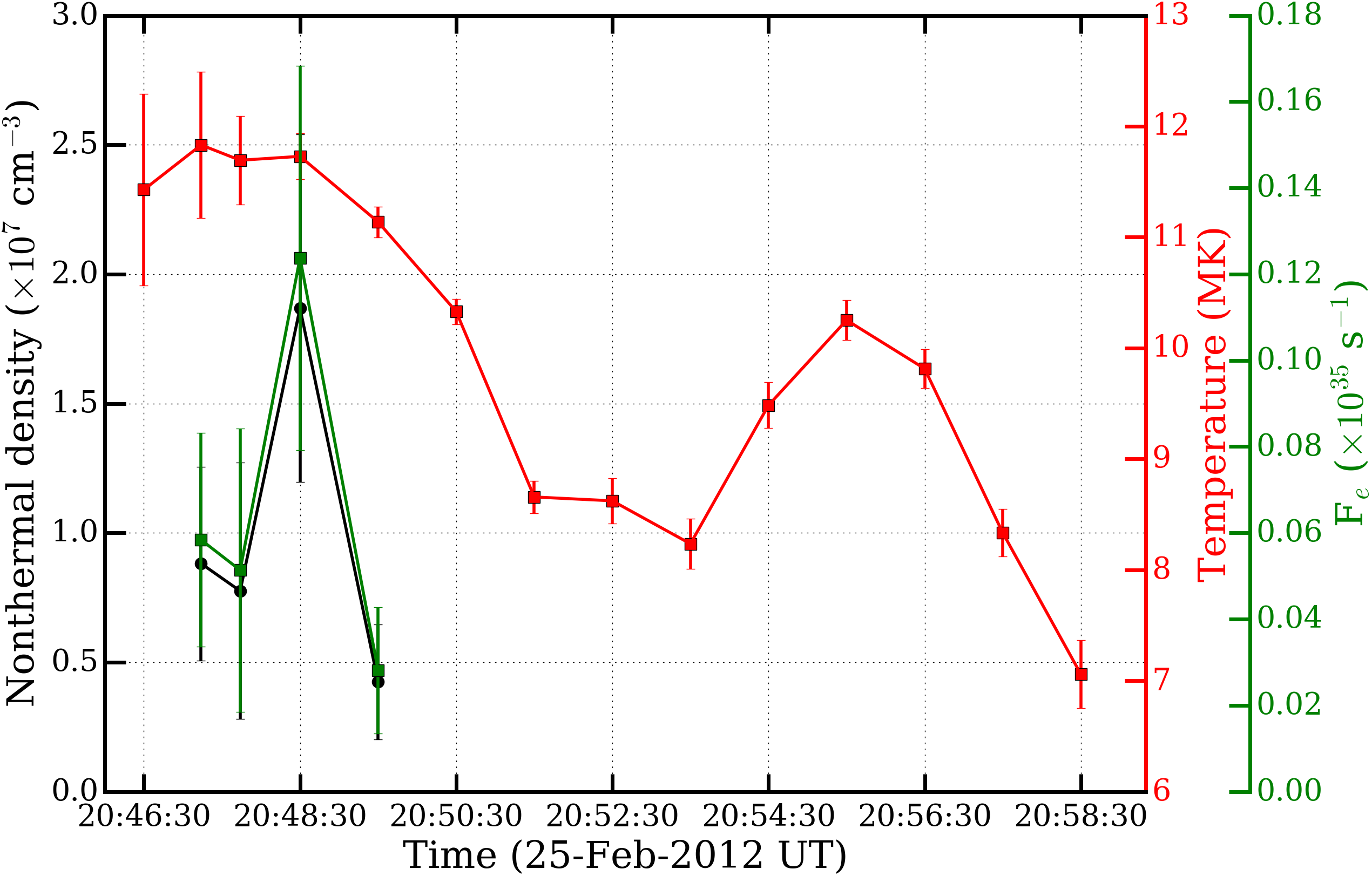}
&
\includegraphics[trim = 0cm 0cm 0.0cm 0.0cm,width=0.5\textwidth]{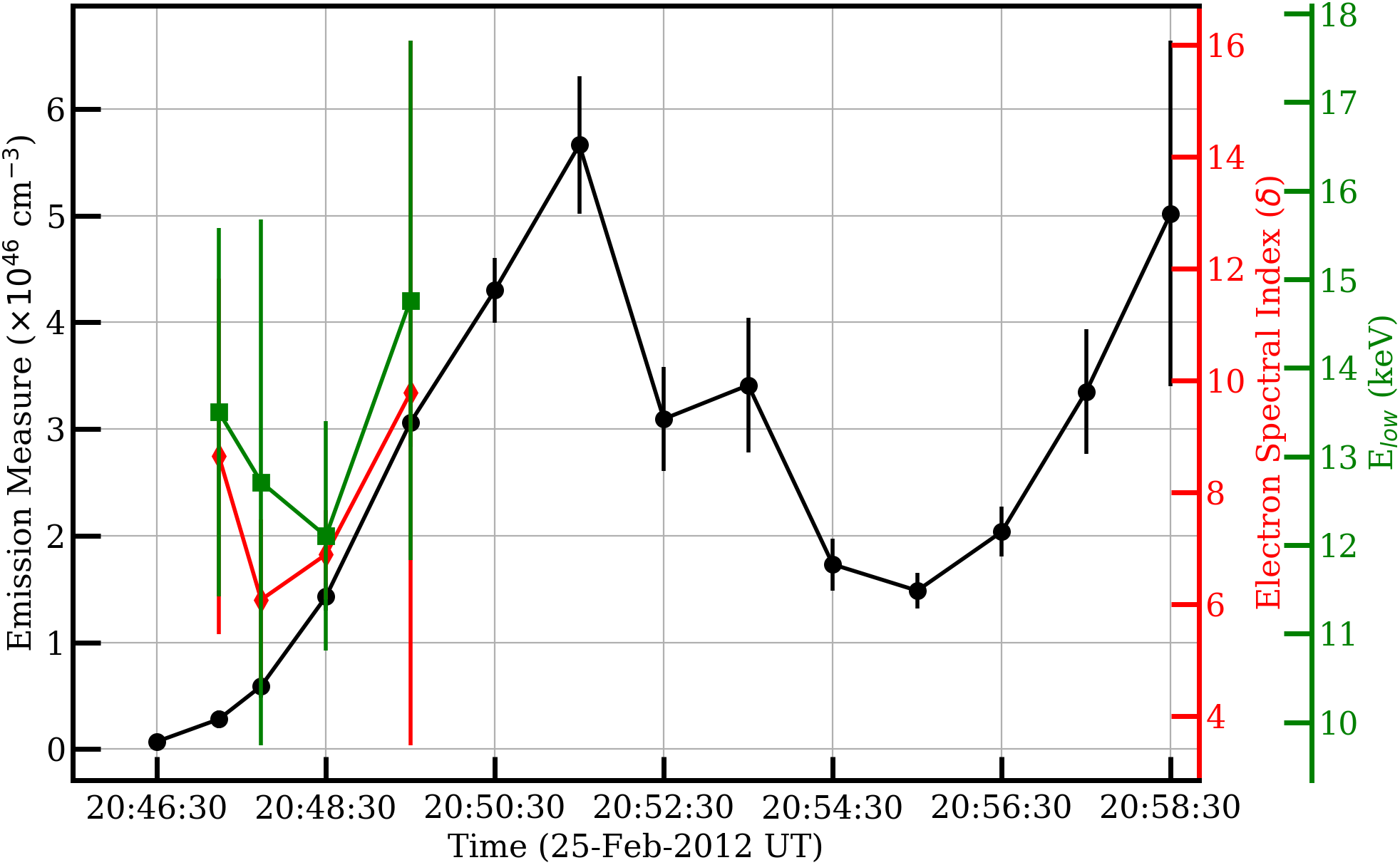}
\end{tabular}
    \caption{Time evolution of flare parameters inferred from RHESSI fitting. Left panel: The nonthermal electron density (black), temperature (red) and total flux (green) variation in time. Right panel: Emission measure (black), electron spectral index (red) and low energy cut-off (green) as a function of time.}
    \label{Fig:rhessi_param}
\end{figure*}

Figure \ref{Fig:rhessi_spec} shows the RHESSI spectrum between 20:47:00 UT and 20:47:28 UT. Note that radio bursts A, B and C occurred during this time. The result of the spectral fit corresponding to a thermal and a nonthermal model is plotted in Figure \ref{Fig:rhessi_spec}. The weak nonthermal component is present in the spectrum up to about 20 keV (above which the background noise dominates the spectrum), with an electron spectral index $\delta$ of $8.6\pm3.2$ and a low-energy cut-off ($E_{\mathrm{low}}$) of 13.5 keV.
Additionally, the nonthermal thick-target fit yields the total electron flux ($F_e$; i.e., total number of nonthermal electrons  above the cut-off energy $E_{\rm low}$ per unit time). \\ 
From the thermal and nonthermal fits, a number of additional plasma parameters, such as nonthermal electron density and thermal background density were deduced as follows. The nonthermal electron density, $n_{\rm nth}$ was calculated as,
 \begin{equation}\label{eq:nthdens}
     n_{\rm nth} = \frac{F_e}{v_e A_{\rm fp}}\quad \mathrm{cm^{-3}},
 \end{equation}
 where $v_e$ is the electron velocity and $A_{\rm fp}$ is the footpoint source area. 
The footpoint area, i.e. the area over which energy is deposited, is difficult to determine from the observations, since source sizes in RHESSI images depend on the imaging algorithm, the specific settings of the chosen algorithm, and the spatial resolution provided by the chosen grids. Hence, we use an estimate of the ribbon area in EUV that is co-spatial with the high-energy X-ray source as a proxy (see 10--18 keV in Figure \ref{Fig:ribbons}). This gives a footpoint area of $3" \times 6" = 18 $ arcsec$^2$, i.e $A_{\mathrm{{fp}}} \approx 9.5 \times 10^{16}$ cm$^2$. The  electron velocity is chosen as $v_e = \sqrt{2E_{\rm low}/m_e}$ = $7 \times 10^{9}$ cm/s. This is the velocity corresponding to an energy of 13.5~keV, the cutoff energy inferred from the RHESSI fit. Since the power-law spectrum is dominated by the lowest energy electrons, this is a good approximation of the average velocity in the distribution. Using the electron flux from the RHESSI fits, $F_e =5.8\times10^{33}\,\mathrm{s^{-1}}$, we obtain an approximate nonthermal electron density $8.8\times 10^{6}$ cm$^{-3}$ above $E_{\rm low} = 13.5$ keV. The background thermal electron density can be found as $n_{\rm th} = \sqrt{EM/V}$, where $EM$ is the volume emission measure from RHESSI spectral fit and $V$ is the source volume. The RHESSI  low-energy source image was used to infer the source area from which the emitting volume was calculated as $V = A^{3/2}=1.3 \times 10^{27}\, \mathrm{cm^{-3}}$, assuming a filling factor of 1. The resulting thermal electron density is $n_{\rm th}=1.5\times 10^{9}$ cm$^{-3}$. Note that this is the density of the thermal source near the top of the flaring loop. The density near the base of the corona, where the high-energy X-ray emission originates, is likely several factors higher.

The temporal evolution of the RHESSI spectral fit parameters is shown in Figure \ref{Fig:rhessi_param}. The plasma temperature shows two distinct peaks. The first peak ($\sim\!12$ MK) coincides with nonthermal emission. The second peak at 20:55:30 UT has no nonthermal counterparts but coincides with a faint brightening in the GOES and RHESSI X-ray flux (c.f., Figure \ref{Fig:overview}), suggesting additional post-flare heating. The nonthermal parameters show a temporal evolution including spectral hardening and softening (red curve in the right hand panel) but with large uncertainties. 

\begin{figure*}[!hbtp]
\begin{center}
\begin{tabular}{ccc}
\includegraphics[trim = 0cm 0cm 0.0cm 0.0cm,width=0.3\textwidth]{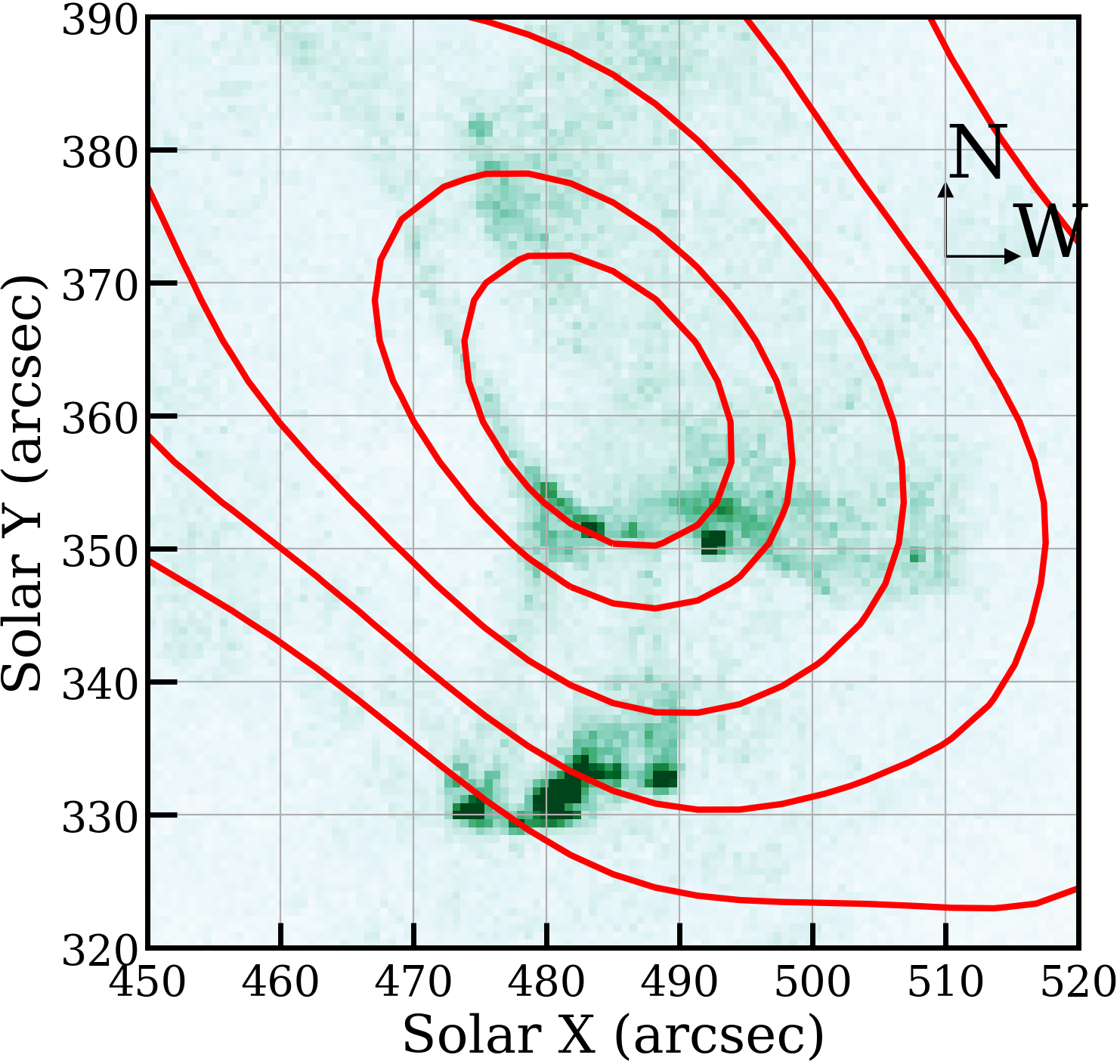} 
&
\includegraphics[trim = 0cm 0cm 0.0cm 0.0cm,width=0.3\textwidth]{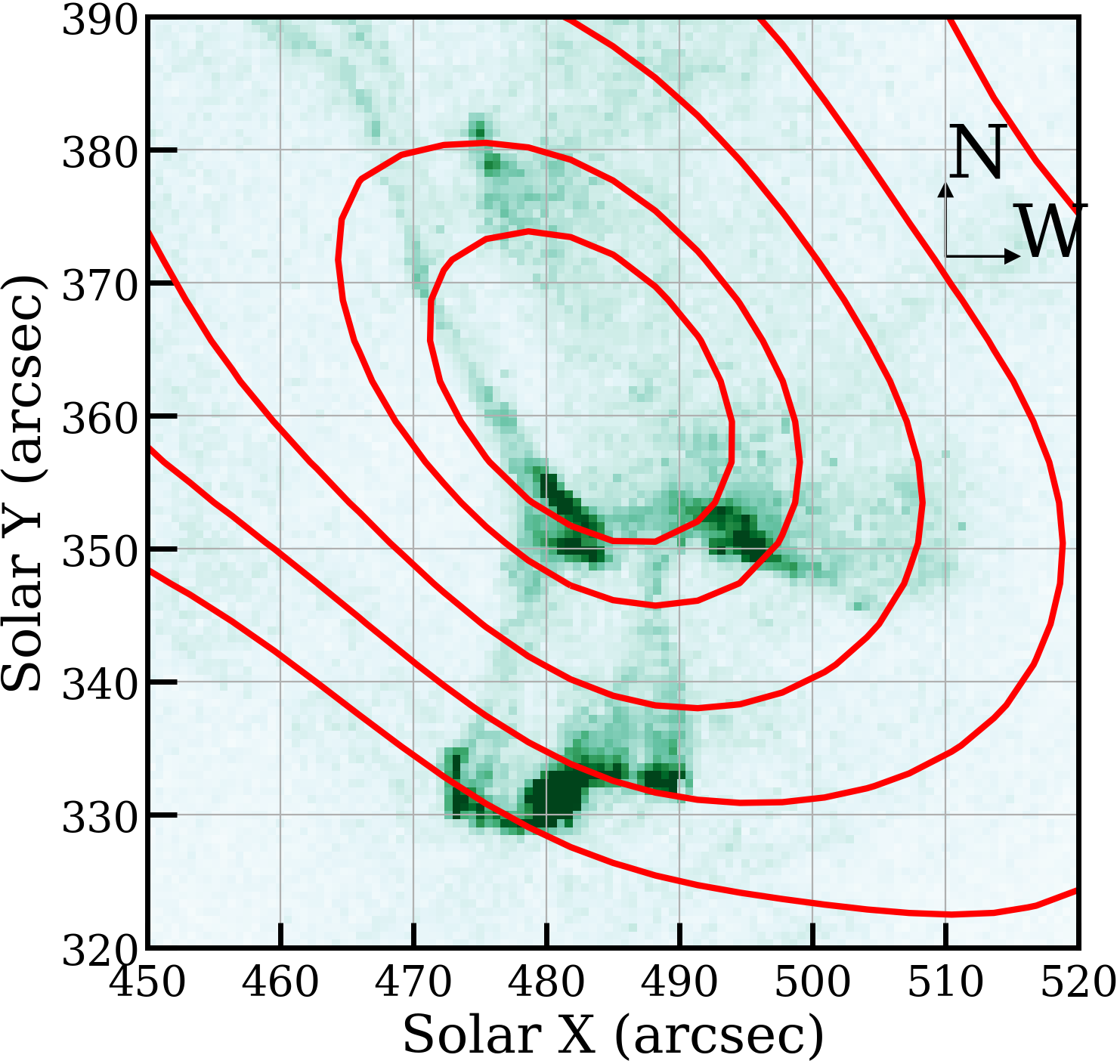}
&
\includegraphics[trim = 0cm 0cm 0.0cm 0.0cm,width=0.3\textwidth]{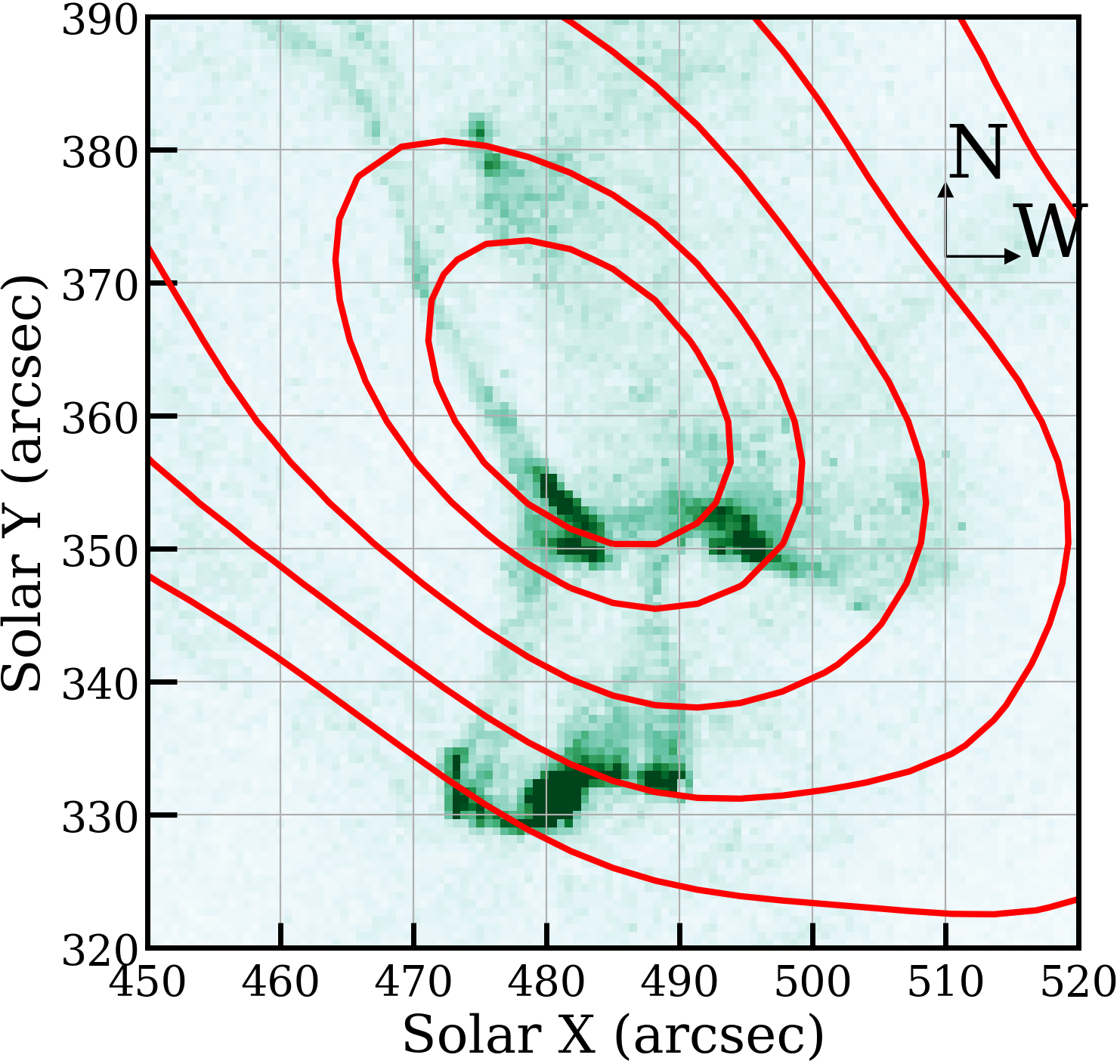}\\
(a) Burst A (20:47:09 UT) & (b) Burst B (20:47:18 UT) & (c) Burst C (20:47:20 UT)
\end{tabular}
\caption{AIA 94 \AA\ EUV ribbons (background image) and VLA radio contours at 2 GHz are shown for three times. The AIA start times are from 20:47:02 UT for burst A and  from 20:47:14 UT for bursts B and C. The contours are at 95, 90, 80, 70, and 60 \% w.r.t the maximum brightness temperature in the corresponding maps.} 
\label{Fig:contours}
\end{center}
\end{figure*}

\subsection{Radio imaging and spectral analysis}
\label{subsec:radio_imaging_analysis}
Radio imaging was performed between 20:40 to 21:00 UT with a 1-s cadence and a frequency average over four 1-MHz-wide spectral channels (4 MHz). Standard radio data reduction and synthesis imaging steps, including flagging, calibration, and deconvolution, were performed. Celestial source 3C48 was used as the flux and bandpass calibrator. Gain calibration was performed against celestial source J2130+0502. For solar scans, 20 dB attenuators are inserted in the signal path to reduce the antenna gain, and corresponding corrections in phase and amplitude were made to the data \citep{2013ApJ...763L..21C}. 
Timeseries radio images suggest the presence of a background continuum, probably associated with the quiescent active region, prior to the multitudes of radio bursts. In order to isolate the radio bursts, we took the average visibilities of a pre-flare time period from 20:46:00 to 20:47:00 UT as the background and subtracted them from the data during the burst periods. The resulting visibilities were used to generate images of the radio bursts. 
Subsequently, the flux calibrated images were converted into brightness temperature maps. The radio images show a dominant source in the observed active region for the duration of the analyzed time interval. The radio source size is several times larger than the synthesised beam, implying that the radio source is extended.
To determine the centroid location, the 90\% radio source contour w.r.t to the maximum in each image was fitted with a 2-D ellipse. Figure \ref{Fig:contours} shows the location of the radio contours of radio bursts A, B and C at 2 GHz.

\subsubsection{Gyrosynchroton fitting} \label{sec:GSfit}

\begin{figure*}
\begin{center}
\includegraphics[trim = 0cm 0cm 0.0cm 0.0cm,width=0.8\textwidth]{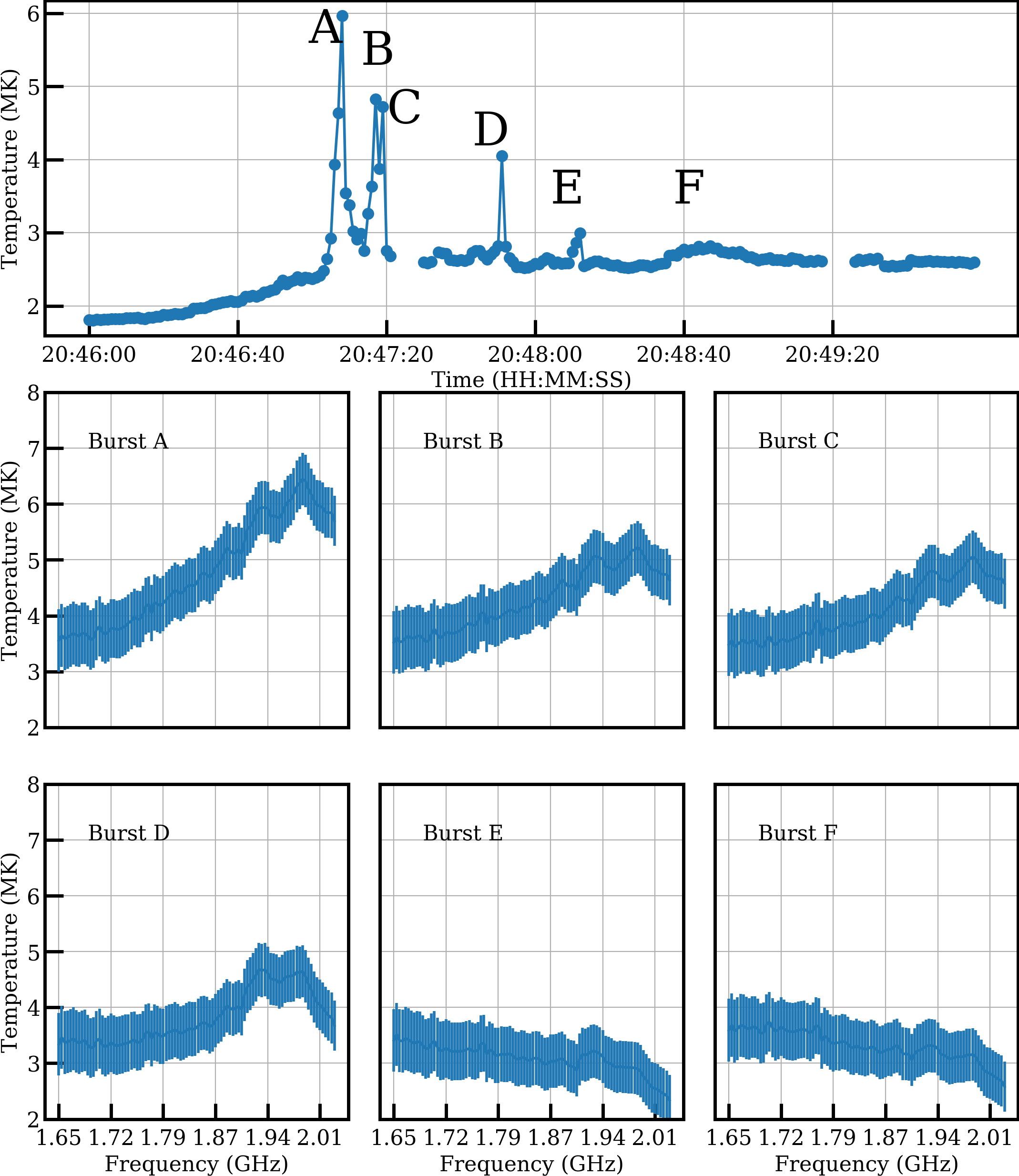}  
\caption{Top panel: Timeseries of the brightness temperature of the radio source obtained from the region corresponding to the 70\% flux level at 2.0 GHz frequency in the image. Middle and lower panels: Radio spectrum for the various 
bursts seen in the radio source timeseries. The errorbars are the 1-$\sigma$ variation in the image residuals.
} \label{Fig:spectrum}
\end{center}
\end{figure*}

The spectra of the radio bursts were extracted in the 1.65 to 2.03 GHz frequency range from 20:46 to 21:00 UT at 4 MHz spectral resolution and 1 second temporal resolution.  The radio source was brighter at higher frequencies and is marginally displaced across 1.65 GHz to 2.03 GHz. Hence for the extraction of the brightness spectrum, we chose a region corresponding to 70\% at 2.0 GHz in the brightness temperature maps and kept this region of interest for all frequencies.
Figure~\ref{Fig:spectrum} shows the brightness temperature spectrum for bursts A to F. The errorbars on the measured spectrum are the 1-$\sigma$ variation of the noise in the image, computed over a region away from the radio source. For most times the brightness temperature spectrum remained flat at about 3--4 MK, suggesting an optically-thick thermal origin. During bursts A to C, however, the spectra developed a positive slope, indicative of optically-thick nonthermal gyrosynchrotron radiation \citep[see, e.g.,][]{Dulk1985}. 

\begin{figure*}
    \begin{center}
    \includegraphics[trim = 0cm 0cm 0.0cm 0.0cm,width=0.9\textwidth]{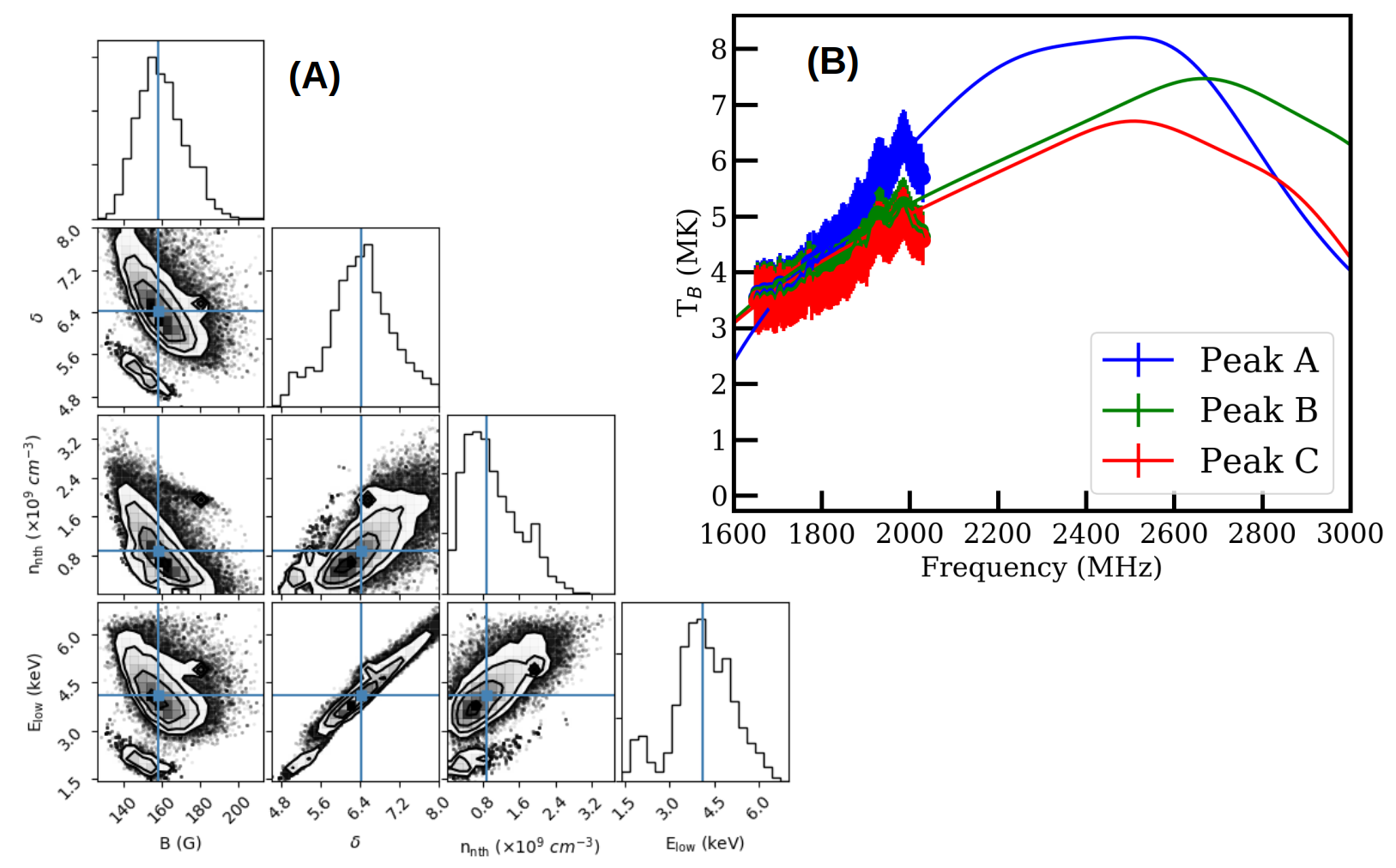}
  \caption{Panel A: The resultant 2-D distribution among various parameters obtained from the MCMC run for the first burst (burst A) are shown in black contours. In each figure the contours represents 86, 68, 39 and 12\% levels w.r.t maximum of the distribution. Individual distribution of each parameter is plotted as the histogram above each parameter column. Panel B: The spectrum of the brightness temperature for the three fitted, individual bursts (labelled A, B, and C, compare Section \ref{sec:GSfit}) are shown in blue, green and red respectively. The solid lines shows the optimum MCMC fit represented in their colors. }
  \label{Fig:mcmc}
  \end{center}
\end{figure*}

Gyrosynchroton emission has been studied in great detail in the context of flares \citep[see, e.g., reviews by][]{Dulk1985,Alissandrakis1986,1998ARA&A..36..131B}. Its characteristic hill-top shape consists of low-frequency and high-frequency power-laws corresponding to the optically thick and optically thin regimes, respectively. The spectral shape and brightness depend on the properties of the accelerated electron distribution and the ambient plasma, including magnetic field strength, nonthermal electron density, and thermal plasma density, among others. Due to this multi-parameter nature of the emission, gyrosynchrotron fitting is a relatively complex problem. 



\begin{table}[]
    \centering
    \begin{tabular}{|c|c|}
    \hline
     RHESSI Parameters & Values  \\
     \hline
        Nonthermal density ($n_{\rm nth}$) & $(8.8\pm 3.7) \times 10^6$ cm$^{-3}$ \\
        Low-energy cut-off ($E_{\rm low}$) & 13.5 $\pm$ 2.0 keV \\
        Thermal density ($n_{\rm th}$) & $(1.5 \pm 0.3)\times10^9$ cm$^{-3}$ \\
        Electron spectral index ($\delta$) & 8.6$\pm$3.2 \\
        Emission Measure ($EM$) & (2.8$\pm$0.8) $\times 10^{45}$ cm$^{-3}$\\
        Temperature ($T_{e}$) & 11.8 $\pm$ 0.6 MK \\
        Electron flux ($F_{e}$) & (5.8 $\pm$ 2.5)$\times10^{33}$ s$^{-1}$ \\
        \hline
    \end{tabular}
    \caption{RHESSI fit parameters closest to the radio bursts of which the spectrum was fitted. The time interval of RHESSI fit is 20:47:00 to 20:47:28 UT.}
    \label{Tab:rhessi_params}
\end{table}


\begin{table*}[]
    \centering
    \begin{tabular}{|c|c|c|c|c|c|c|c|c|}
    \hline
    Burst & $B$ & $n_{\rm nth}$  &  $E_{\rm low}$  & $\delta '$ &$\delta$&$n_{th}$ (fixed) &$F_e$ & Total Power $P$ \\
    &(Gauss)&($\times$10$^8$ cm$^{-3}$)&(keV)& & &($10^{10}cm^{-3}$)&($10^{33}$electrons s$^{-1}$)&($10^{27}$erg s$^{-1}$)\\
    \hline
    A & 158.6$\pm$12.6 & 8.8$\pm$6.4 & 4.1$\pm$0.9 & 6.4$\pm$0.6 & 5.9$\pm$0.6&2.0&320$\pm$230&2.6$\pm$1.9 \\
    B & 176.7$\pm$35.3&16$\pm$11&4.5$\pm$1.4&7.6$\pm$1.4 &7.1$\pm$1.4&1.5&600$\pm$410&5.2$\pm$3.6\\
    C & 170.4$\pm$31.3& 13$\pm$10 & 4.2$\pm$1.1 & 7.6$\pm$1.3 &7.1$\pm$1.3 &1.5& 470$\pm$360&3.8$\pm$2.9\\
    \hline
    RHESSI & - & $0.088 \pm 0.37$ & $13.5 \pm 2.0$ & 9.1$\pm$3.2&8.6$\pm$3.2&$-$ & 5.8 $\pm$ 2.5 &0.15$\pm$0.06\\
    \hline
    \end{tabular}
    \caption{Electron spectral parameters from gyrosynchrotron fitting and the corresponding values from X-ray spectral fitting.}
    \label{Tab:GSfit}
    \tablecomments{$\delta '$ denotes the electron spectral index of the differential density spectrum found by gyrosynchrotron fitting. The spectral index $\delta$ of the differential flux spectrum, found by the thick-target X-ray spectral fit, is related to $\delta '$ as $\delta=\delta '-0.5$. \citep[see][]{2018SSRv..214...82O}. For easier comparison, both values are given in the table for bursts A to C. }
\end{table*}

For calculating the model gyrosynchrotron spectrum, the user-friendly and computationally inexpensive fast GS code \citep{Fleishman2010} was used. In addition to realistic starting parameters, the optimization must take into account the errors on the observed brightness spectrum and return the uncertainties on the optimised parameters. We investigated many different optimization algorithms like Basin-hopping, Markov chain Monte-Carlo (MCMC), Least-Squares and Brute-force methods (detailed explanation in Appendix A). 
Among them, MCMC provides optimal estimates of uncertainties and hence is best suited for the present task. We adopted the MCMC method described in \citet{2020NatAs.tmp..150C} for the spectral fitting. The open-source Python package {\tt emcee} \citep{2013PASP..125..306F} included in the minimization package {\tt LMFIT} \citep{lmfit} was used to sample the posterior probability distributions (PPDs) of the fit parameters based on Bayesian statistics \citep{2010CAMCS...5...65G}. 

\begin{figure*}
\begin{center}
\begin{tabular}{ccc}
\includegraphics[trim = 0cm 0cm 0.0cm 0.0cm,width=0.33\textwidth]{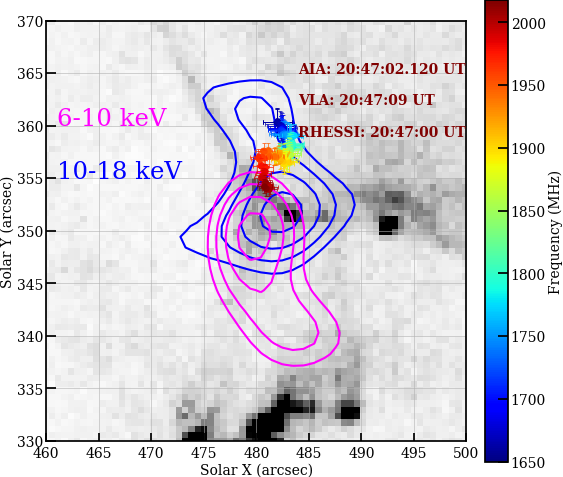} 
&
\includegraphics[trim = 0cm 0cm 0.0cm 0.0cm,width=0.33\textwidth]{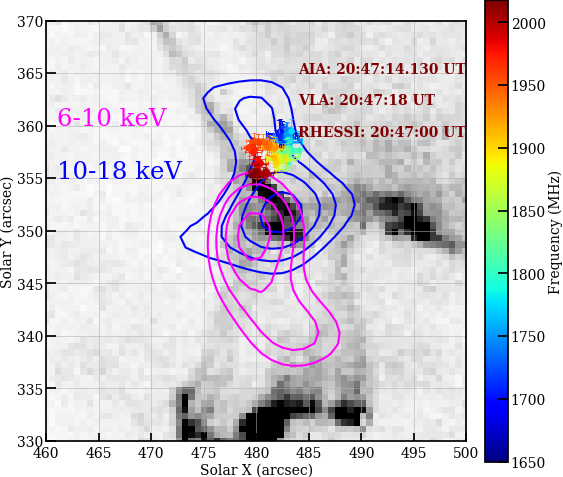}
&
\includegraphics[trim = 0cm 0cm 0.0cm 0.0cm,width=0.33\textwidth]{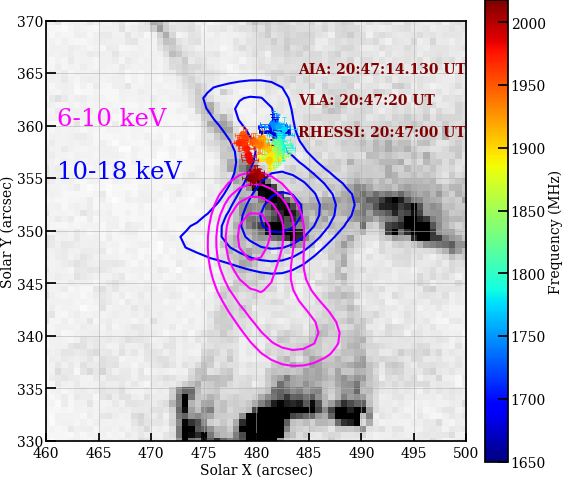}\\
(A) & (B) & (C)\\
\end{tabular}
\begin{tabular}{cc}
\includegraphics[trim = 0cm 0cm 0.0cm 0.0cm,width=0.33\textwidth]{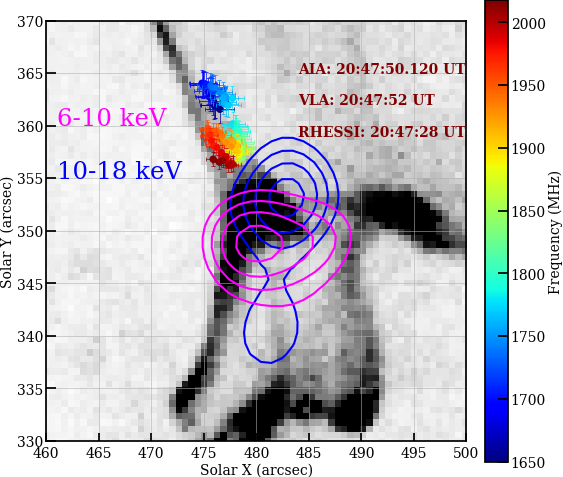}
&
\includegraphics[trim = 0cm 0cm 0.0cm 0.0cm,width=0.33\textwidth]{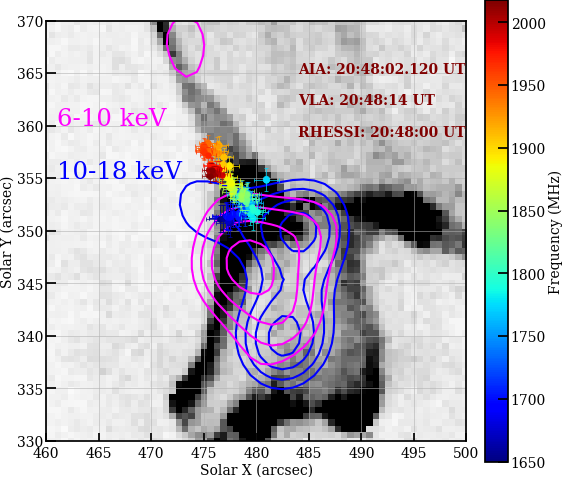}
\\ 
(D)&(E)\\
\end{tabular}
\caption{Evolution of the AIA 94 \AA \ EUV ribbons (black and white image, color table inverted), X-ray RHESSI sources (magenta and blue contours) and VLA radio centroid (crosses) positions. The RHESSI contour levels are at 65\%, 75\%, 85\% and 95\% w.r.t map's peak. The AIA and RHESSI images are temporally closest to the radio bursts. The start times of AIA, VLA and RHESSI images are displayed in each panel. Note that RHESSI has an integration time of 28 sec in A, B and C, but 1 minute for D and E. The colorbar indicates the frequencies of plotted radio centroids.} 
\label{Fig:image_all}
\end{center}
\end{figure*}

Here, we assumed a homogeneous nonthermal source. We treated magnetic field strength ($B$), nonthermal density ($n_{\rm nth}$), thermal density ($n_{th}$), electron spectral index ($\delta '$), and low-energy cut-off ($E_{\rm low}$) as free parameters. We fixed all other parameters either using typical values from the literature or inputs from additional observational constraints. Specifically, column depth and viewing angle were chosen to be $10^{9}$ cm and 60$^o$ respectively. 
The initial values of the five free parameters were constrained based on the HMI magnetogram and the X-ray fitting results.
We note that with thermal density as a free parameter, the MCMC process did not converged. Hence, the thermal density was determined by manually investigating the parameter space and fixing it to $n_{th}=2\times10^{10}\, \mathrm{cm}^{-3}$,  $1.5\times10^{10}\, \mathrm{cm}^{-3}$ and  $1.5\times10^{10}\, \mathrm{cm}^{-3}$ for bursts A, B and C respectively.
Figure \ref{Fig:mcmc} shows a ``corner plot'' of the resulting PPDs from the MCMC optimization for the free fit parameters and the resulting model-spectra for burst A. Here, the top 1-D histogram shows the MCMC PPD of each parameter. The 2-D histogram shows the PPD for a pair of parameters, i.e. showing six 2D-distributions for the four parameters. The MCMC output shows bell-shaped distributions for each parameter, i.e. the parameter distributions are well constrained. Such 2-D representation of the PPDs is useful for visualising trends and correlations among parameters.  
In the present case, a weak anti-correlation of magnetic field strength $B$ with $E_{\rm low}$, $n_{\rm nth}$, and $\delta '$ can be seen. All other parameters, i.e. $\delta '$, $n_{\rm nth}$ and $E_{\rm low}$ are positively correlated with each-other.
However, one must be careful as some correlations can be due to the interdependence between pairs of the fit parameters \citep[see, e.g., Movie S2 in][]{Fleishman2020}. For example, both the magnetic field and the non-thermal density affect the peak frequency of the spectrum. An increase in magnetic field strength results in a higher peak frequency. The same effect can be achieved by increasing the nonthermal density. Hence, magnetic field strength and nonthermal density are expected to be anti-correlated in the fitting results, as can be seen in Figure \ref{Fig:mcmc}.

Table \ref{Tab:GSfit} lists the optimised values and corresponding uncertainties of the parameters. 
We note that the low-energy cutoff of the nonthermal electron distribution $E_{\rm low}$ is smaller than that inferred from the RHESSI X-ray spectrum by a factor of $\sim$4, while the nonthermal density from the gyrosynchrotron fit is two orders of magnitude higher than the RHESSI estimates. This discrepancy in $E_{\rm low}$ and $n_{\rm nth}$ is present in burst A through C, a possible indication that the two instruments observe two different electron populations.
Table~\ref{Tab:GSfit} also lists the uncertainty in each parameter. We note that these are large, especially for bursts B and C. 
The dominant reasons for the large uncertainties are the limited frequency range available for gyrosynchroton fitting, and the Signal-to-Noise Ratio (SNR) of the data. Due to the limited frequency range, neither the peak of the spectrum nor the optically thin part were observed. Hence, both, the magnetic field strength and the spectral index are not well constrained. The rather low SNR resulted in relatively large uncertainties of the observed spectra and consequently large uncertainties of the fitted parameters.
The SNR of burst A is better than burst B and C, i.e. the spectral fit is relatively well-constrained, and the fitted parameters are more reliable. 

\section{Spatial evolution of flare parameters}\label{s:evol}



\begin{figure}
\begin{center}
\includegraphics[trim = 0cm 0cm 0cm 0cm,width=0.5\textwidth]{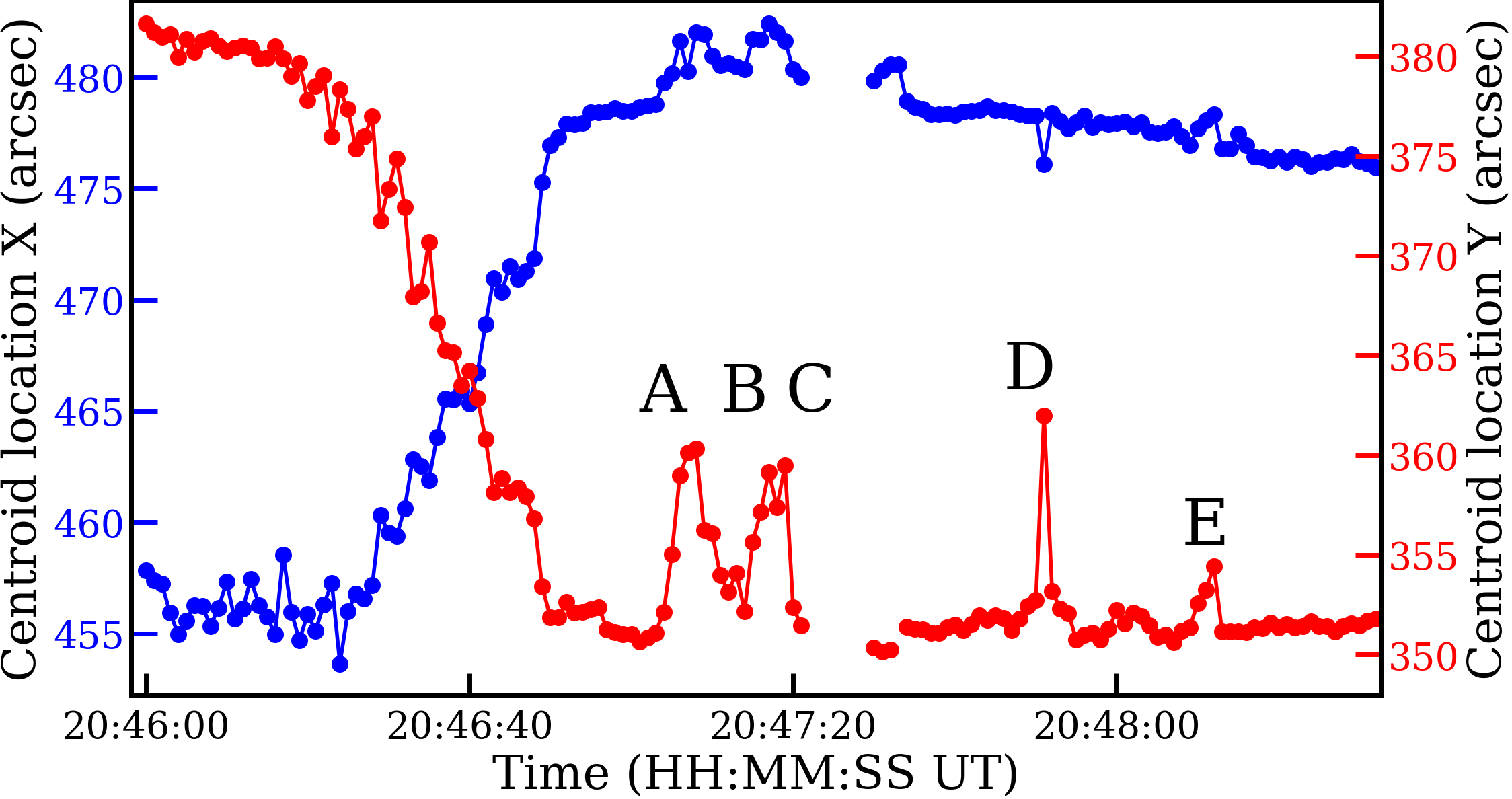}  
\caption{Centroid locations of the radio sources as function of time at 1.7 GHz. The position shifts during the burst times can be seen distinctly and are marked with letters. Note: X and Y coordinates correspond to the helioprojective solar coordinates shown in Figure \ref{Fig:image_all}.
} \label{Fig:centroids}
\end{center}
\end{figure}

The spatial evolution of the radio bursts A, B and C shows interesting behaviour that is discussed in more detail in the following. 
Before the main bursts, the centroid locations show a large scatter. During the radio bursts, the source centroids at all frequencies become clustered together within 7" near the northern ribbon (e.g. Figure \ref{Fig:image_all} A). 
The observed compact clustering suggests a bright common radio source for all shown radio frequencies.  Figure \ref{Fig:centroids} shows the temporal variation of the radio centroid locations at a representative frequency of 1.7 GHz. The radio centroid location varies during the flare. A distinct change in the centroid location of the radio source is seen during each burst.
Since the observed displacement in the centroids in Figure \ref{Fig:centroids} is co-temporal with the brightness temperatures of the bursts (Figure \ref{Fig:spectrum}, top panel), these displacements are significant and real.
We also computed positional uncertainties ($\sigma_{X,Y}$) for the radio sources by ${\sigma_{X,Y} = \frac{\theta_{FWHM}}{SNR \sqrt{8 ln(2)}}}$, where $\theta_{FWHM}$ is the FHNW of the synthesised beam and SNR is the signal-to-noise of the radio source \citep{2015Sci...350.1238C,2018ApJ...866...62C}. However, these uncertainties can be larger as the radio source is extended. 

For bursts A, B and C, the change in the centroid locations are most pronounced (see peaks A, B \& C in Figure \ref{Fig:centroids}). Here, the heights of the peaks indicate a displacement of $5"$ and $10"$ in solar X-coordinate and Y-coordinate, respectively. Interestingly, the X and Y coordinates for bursts A, B, and C are similar, suggesting a single burst location. 
During this time, both low-energy and high-energy X-rays show significant emission (Figure \ref{Fig:image_all} A, B \& C). Both X-ray sources and radio bursts A, B and C appear near the northern ribbon, but they are not co-spatial with each-other. The location of these burst sources suggests a low-coronal origin of the emission (see Section \ref{s:discuss}). We note that the low and high energy X-ray sources show an extension at 65\% and 75\% contours towards southern and northern ribbon, respectively (Figure \ref{Fig:image_all} (C)).

The location of bursts D and E is different from bursts A, B, and C (Figure \ref{Fig:centroids}). The brightness temperature of burst F is too low for determining centroid positions accurately. Burst D and corresponding low and high energy X-ray sources lie on the northern ribbon (Figure \ref{Fig:image_all} (C)). Like for bursts A, B and C, the X-ray sources, and radio bursts are not co-spatial. During burst E, the high-energy X-ray source shows an extension toward the southern ribbon, while the low-energy X-ray source and the radio bursts are located near the northern ribbon. However, the high energy X-ray source for burst E is faint, and its significance is relatively small. 

\begin{figure*}
    \centering
    \begin{center}
    \begin{tabular}{cc}
       \resizebox{52mm}{!}{
\includegraphics[trim={0.0cm 0cm 0.0cm 0.0cm},clip,scale=0.13]{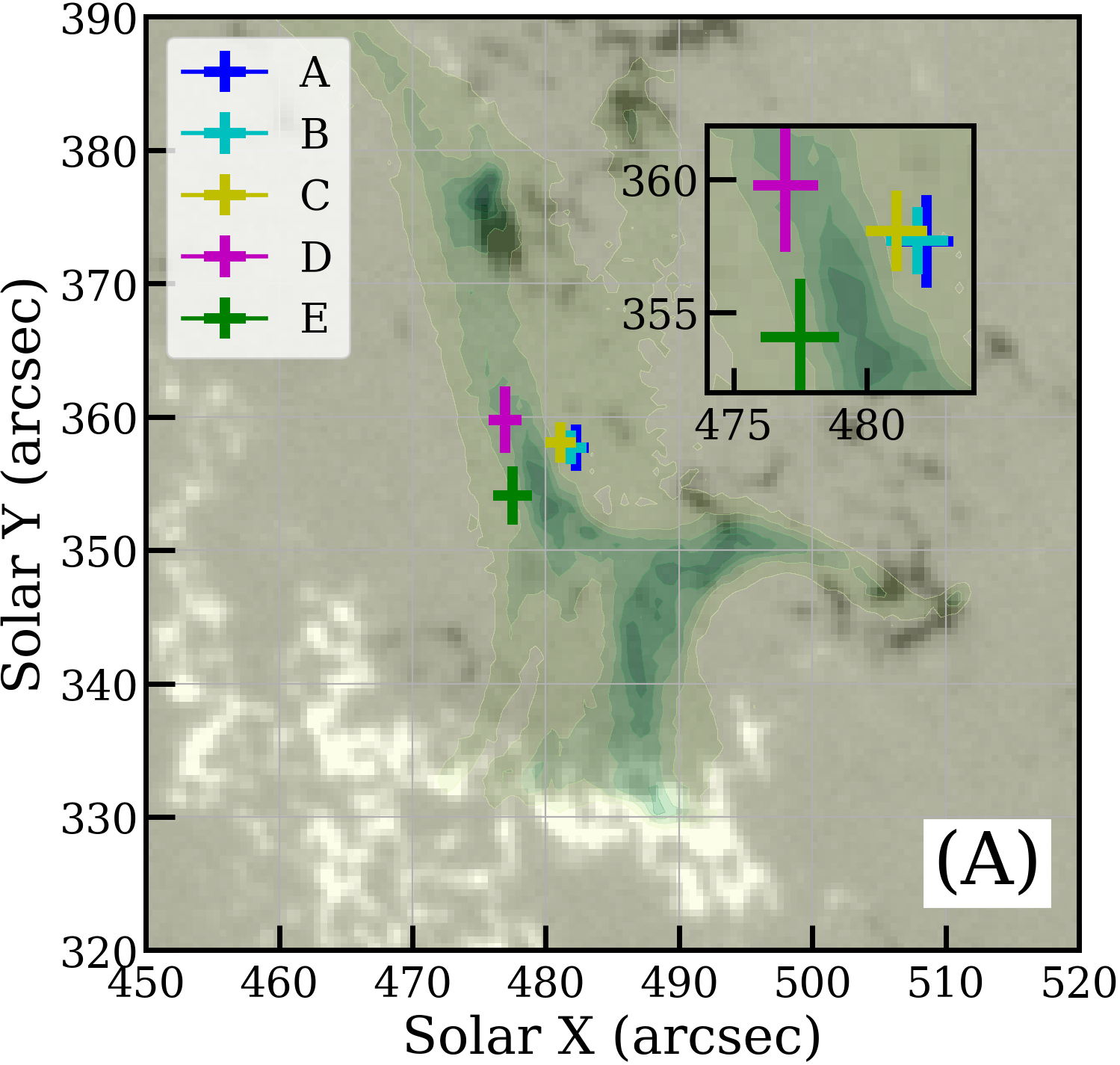}}
        &  
        \resizebox{67mm}{!}{
\includegraphics[trim={0.0cm 0cm 0.0cm 0.0cm},clip,scale=0.13]{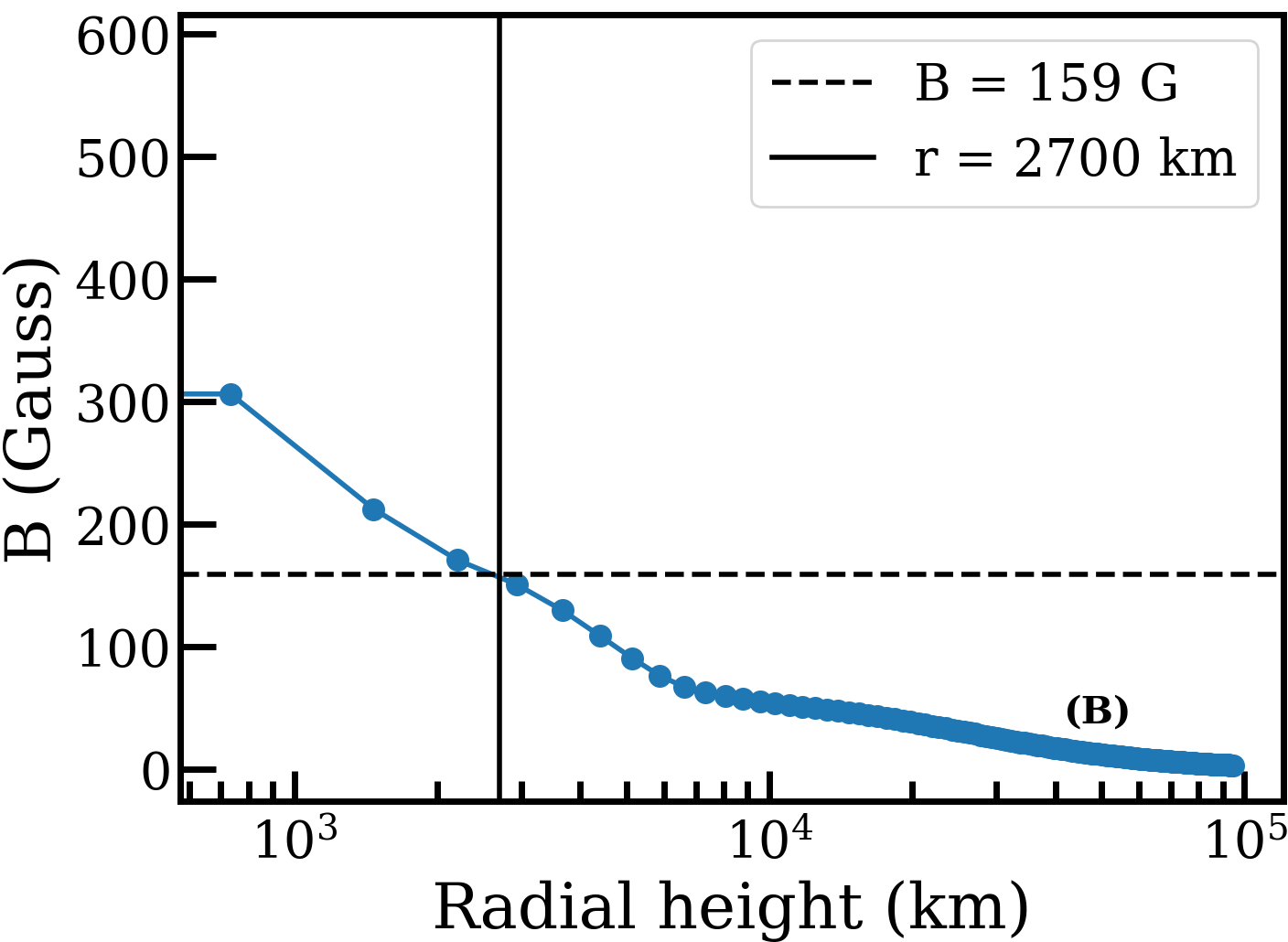}}
        \\
    \end{tabular}
        \end{center}
      \begin{center}  
    \begin{tabular}{ccc}
         \resizebox{55mm}{!}{
\includegraphics[trim={0.0cm 0cm 0.0cm 0.0cm},clip,scale=0.13]{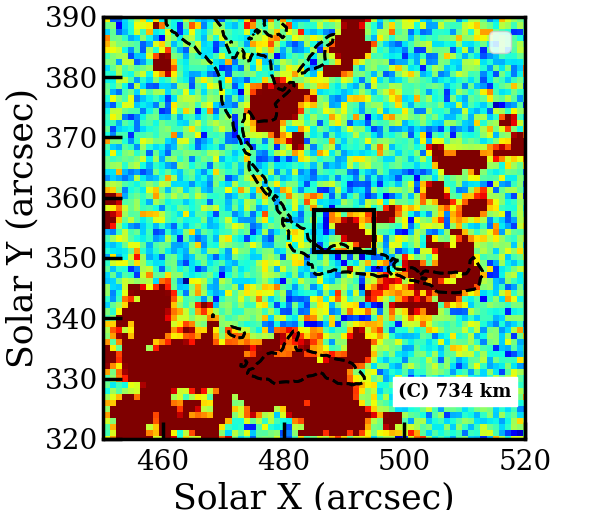}}
         &  
\resizebox{54mm}{!}{
\includegraphics[trim={0.0cm 0cm 0.0cm 0.0cm},clip,scale=0.13]{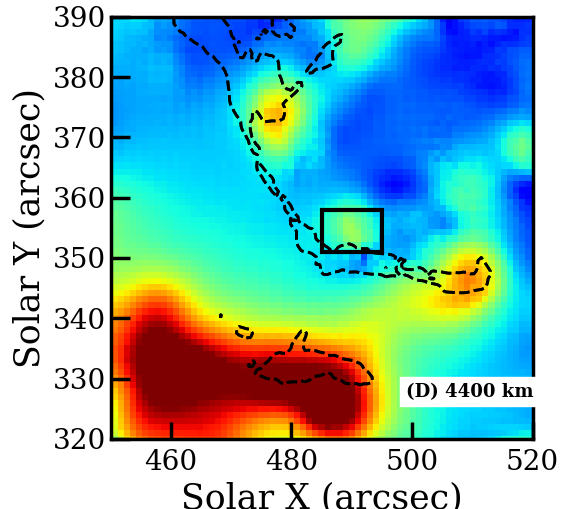}}
    & \resizebox{58mm}{!}{
\includegraphics[trim={0.0cm 0cm 0.0cm 0.0cm},clip,scale=0.13]{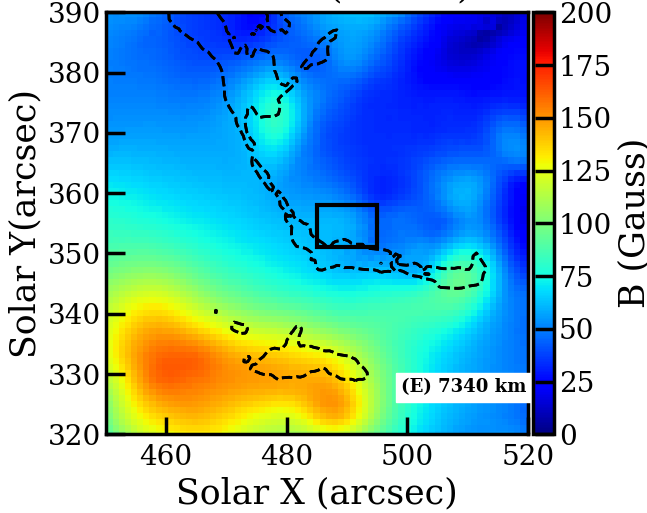}} \\
    \end{tabular}
        \end{center}
    \caption{ Panel A: Composite map of HMI magnetogram, AIA 94 \AA\ and radio centroids of the bursts. The background black and white image is HMI magnetogram from 20:50:43 UT. The overlayed semi-transparent green image shows the AIA 94 \AA\ map at radio burst F time. The symbol of plus marks the centroid positions of bursts from A to E. Their lengths and widths represents the 1-$\sigma$ variation of centroid positions in frequency from 1.65 to 2.03 GHz. Panel A inset: Zoomed-in map of left panel showing variation in the radio centroid locations. Panel B: Radial profile of the maximum magnetic field in the shown rectangular region in panel C, D and E. Here dashed and solid grey lines shows the fitted magnetic field value and corresponding coronal height respectively for burst A. 
    Panel C, D and E: Snapshots from the extrapolated 3D magnetic field configuration at different heights. The dashed-black curve shows the location of the ribbons observed in 304 \AA. The solid-black rectangle marks the closest high magnetic field region ($>200$ G) to the radio sources.}
    \label{Fig:position_final}
\end{figure*}

To further investigate the spatial origin of the radio source, we perform 3D coronal magnetic field extrapolation provided by the {\tt gx\_simulator} package (part of the IDL {\tt SolarSoftware} distribution; \citealt{Nita2015}) and compare with the magnetic field strength inferred from gyrosynchrotron fitting results. Figure \ref{Fig:position_final} (B) shows three solar-disk projected maps of coronal magnetic field strength obtained at three horizontal slices made in the 3D extrapolation at 734 km, 4,020 km, and 7,340 km above the photosphere. A strong asymmetry in magnetic field strength between the northern and southern ribbon can be observed, with the magnetic field strength in the northern ribbon decreasing more rapidly as a function of height. This can be seen in Figure \ref{Fig:position_final} (C, D and E), which shows the maximum magnetic field strength (within a solid-black rectangle) near the northern ribbon at different heights. According to this, the height corresponding to the magnetic field value from the gyrosynchrotron fit (159 G) is 2,700 km. 




\section{Discussion}\label{s:discuss}
The observations presented here provide a multi-wavelength picture of multiple acceleration events during the early phases of a solar microflare at a temporal resolution of 1 s in radio wavelengths, co-temporal with high-energy X-ray sources. Microwave observations allowed us to infer the spectral properties of the accelerated electron distribution, as well as the ambient magnetic field strength. The two main outcomes of this study are: 
\begin{enumerate}
    \item The radio emission originates from the low corona, close to the footpoint of the magnetic loop
    \item The observed radio and X-ray sources are signatures of entirely different electron populations that were accelerated in different acceleration episodes, potentially at different locations.
\end{enumerate}

 \subsection{Low-coronal origin of radio emission}
In most solar flare observations, the low-energy thermal X-ray sources originate from the top of a flaring loop, or sometimes outline the whole loop, while high-energy X-ray emission is typically from the footpoints of flare loops resulting from the bulk of precipitating nonthermal electrons losing their energy in the chromosphere. Since gyrosynchrotron emission is heavily dependent on the magnetic field strength, emission at microwaves is often due to trapped particles in magnetic loops, making the interpretation of the spatial origin of the emission less clear. In the present case, the radio sources during bursts A to C lie on the northern ribbon (as seen in EUV) located close to, but not exactly co-spatial with the high-energy X-ray footpoint. As demonstrated in Section 4 and Figure 10, magnetic field extrapolation implies a low-coronal origin of the radio sources (altitude of 2,700 km), consistent with their projected location near the flare ribbon. Therefore, the radio burst sources (A, B and C), much like the high-energy X-ray sources, are likely signatures of nonthermal electrons in the low corona.

 \begin{figure*}
\begin{center}
\includegraphics[]{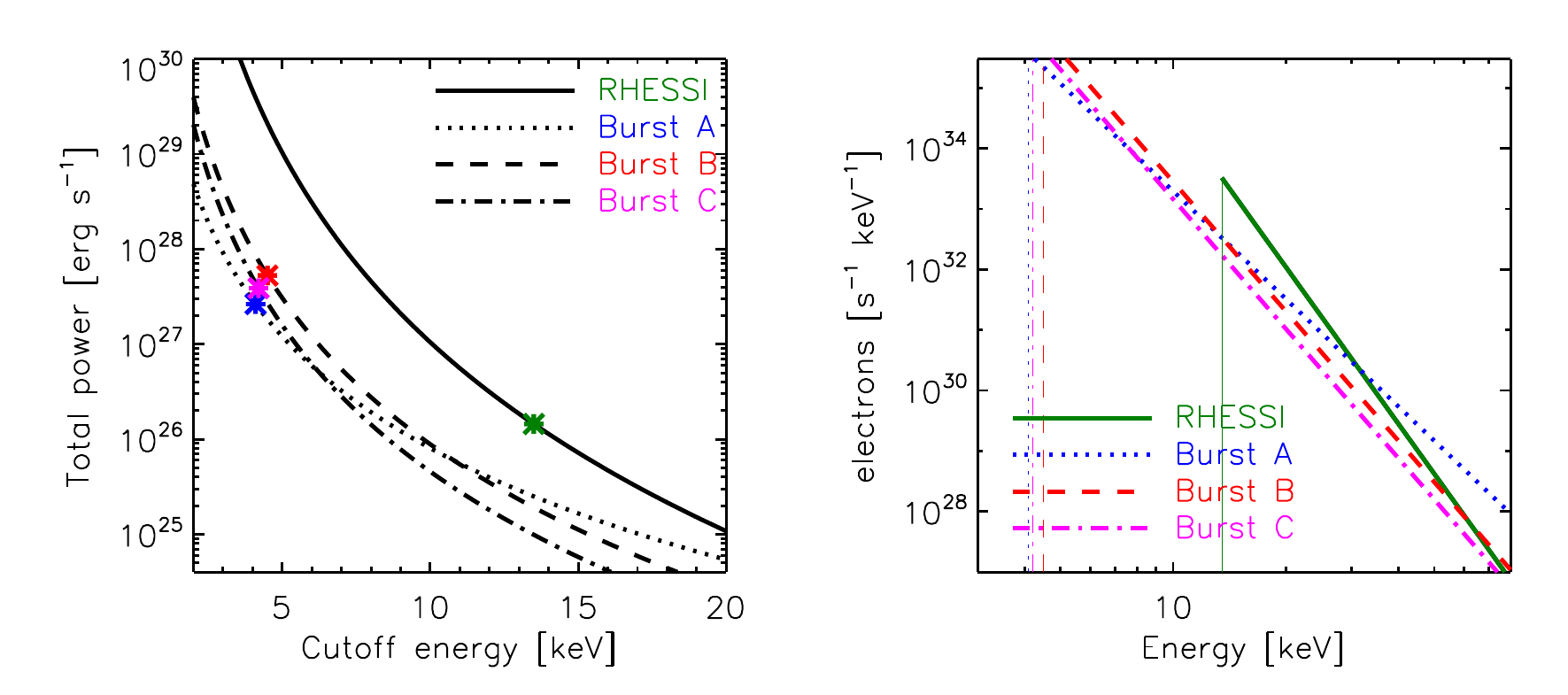}  
\caption{Left: Nonthermal power as a function of low-energy cutoff for the spectral parameters inferred from the X-ray fit (RHESSI) and from the gyrosynchrotron fits of bursts A to C. The symbols mark the nonthermal power calculated from the observed low-energy cutoff. Right: model electron spectra inferred from X-rays (RHESSI) and radio bursts A to C. The vertical lines give the position of the low-energy cutoff. }.
\label{Fig:comppower}
\end{center}
\end{figure*}

\subsection{Multiple acceleration episodes and different electron populatioins}\label{sec:rhessi}
The observations presented here lend strong support to a scenario with multiple acceleration events, possibly in different magnetic loops since, even though the X-ray emission was observed co-temporally with the radio bursts, the observed spectra are very different.  \\ 
Two factors contribute to this interpretation:
\begin{enumerate}
\item The spectral parameters inferred from the X-ray observations are different from the properties inferred from the radio observations to an extent that cannot be explained by uncertainties or by the fact that the X-ray spectrum was time-integrated over the 28 seconds during which bursts A to C were observed.
\item Both, X-ray and radio sources show a remarkable footpoint asymmetry that can only be explained by different electron distributions in different magnetic loops. These two points are elaborated on in the following.
\end{enumerate}
\subsubsection{Different spectral properties inferred from RHESSI and VLA}
The RHESSI X-ray and VLA gyrosynchrotron fits provide the properties of the accelerated electron spectra, like spectral index, total electron flux, and low-energy cutoff. In addition, it is straightforward to calculate the total power in the accelerated electrons from the fitted parameters. For a given spectral index $\delta$, total electron flux $F_e$ (s$^{-1}$), and cutoff energy $E_{\rm low}$ in erg, the nonthermal power can be found as:
\begin{equation}
P=\frac{\delta-1}{\delta-2}F_eE_{\rm low}
\end{equation}
Since the gyrosynchrotron fit returns total nonthermal electron density in cm$^{-3}$, the total electron flux has to be calculated first, using Equation~\ref{eq:nthdens}. Note that the gyrosynchrotron fit gives the spectral index $\delta '$ of the electron differential density spectrum $N(E)$, while the X-ray spectral fit gives the spectral index $\delta$ of the differential flux spectrum $F(E)$. The two indices are related via $\delta=\delta '-0.5$ \citep[see][]{2018SSRv..214...82O}.  As with X-rays, the footpoint area over which energy is deposited is determined from the EUV ribbon area. This estimate results in about the same area as the high energy X-ray footpoint area. With this value, the resulting total powers inferred from the gyrosynchrotron fit are $(2.6\pm 1.9)\times 10^{27}$erg/s, $(5.2\pm3.6)\times 10^{27}$erg/s and $(3.8\pm 2.9)\times 10^{27}$erg/s for bursts A, B and C respectively. 
In comparison, the nonthermal power inferred from the X-ray fit is $(1.5\pm0.6)\times 10^{26}$erg/s, i.e. a factor of ten lower. 
Since the nonthermal power depends on the low-energy cutoff, spectral index, and total electron flux, this discrepancy is unsurprising. While the fitted spectral indices agree, at least within the large uncertainties, the low-energy cutoff found from the gyrosynchrotron fit (around 4 keV) is notably lower than the 13.5 keV found from the X-ray fit. At the same time, the total electron density and, as a result, the total electron flux from the gyrosynchrotron fit, is a factor of 10 to 100 higher than from the X-ray fit. Here, the low-energy cutoff plays a crucial role, since both, the total electron flux and the nonthermal power strongly depend on it, i.e. a higher low-energy cutoff will result in a smaller nonthermal electron density. At the same time, $E_{\mathrm {low}}$ is notoriously difficult to constrain, at least through X-ray spectral fitting. Since the low-energy X-ray emission is dominated by thermal emission, often it is only possible to infer an upper limit of $E_{\mathrm{low}}$ \citep[e.g.][]{2011SSRv..159..107H}. However, as can be seen in Figure~\ref{Fig:rhessi_spec}, the nonthermal photon spectrum clearly turns over, a strong indication that the low-energy cutoff is in fact observed in the present case. Hence, the values of $E_{\mathrm {low}}$ obtained from the gyrosynchrotron fit cannot be seen as potential "true" cutoff energies of the same electron population, but rather the gyrosynchrotron spectra have to be viewed as spectra of a different electron distribution. This point is demonstrated in Figure~\ref{Fig:comppower} where a closer investigation of the total nonthermal powers as a function of low-energy cutoff was made. We calculated the expected nonthermal power from the X-ray spectrum as a function of low-energy cutoff and compared it with the nonthermal power from the gyrosynchrotron spectrum. If the observed X-ray spectrum originated from the same electron distribution as the gyrosynchrotron spectrum but with a too high inferred low-energy cutoff, then the total powers should display a better agreement when calculated using the same low-energy cutoffs. However, this is not the case. The calculated power from the X-ray spectrum for a low-energy cutoff of 3 keV is orders of magnitude higher than from the gyrosynchrotron fit and inconsistent with typical values found for C-class flares. On the other hand, the expected total power in the gyrosynchrotron spectrum for a cutoff energy of $E_{\mathrm {low}}=13.5$ keV is much lower than the calculated power from the X-ray spectrum. In addition to these curves, the model electron spectra are also shown in Figure~\ref{Fig:comppower} with their respective cutoff energies. This illustrates the effect that the cutoff energy has on the total electron flux and hence the inferred power.

\subsubsection{Different spatial origin and asymmetry of footpoint emission}
Another strong indication that RHESSI did not observe the same electron population as the VLA is the location of the sources. As shown in Figure \ref{Fig:image_all}, the source locations of the radio and X-ray sources do not coincide. An even stronger case for this interpretation can be built based on the observed strong footpoint asymmetry, i.e. the complete absence of radio emission as well as a high-energy X-ray footpoint emission at the southern ribbon. While the low-energy, thermal X-ray source clearly lies near the top of a loop arcade (also seen at EUV wavelengths), the nonthermal signatures originate entirely from the northern ribbon. This phenomenon is in contradiction with the standard flare scenario where footpoint emission is expected to present at both ends of the flare arcade. Such footpoint asymmetries have been studied extensively with RHESSI \citep[e.g.][]{2012ApJ...756...42Y,2008SoPh..250...53S,2009ApJ...693..847L}. The most common explanation for such asymmetries is an asymmetry in the magnetic field strength between both legs of the loop, resulting in different mirroring ratios. In such a scenario, the high-energy X-ray emission from the loop end with the higher magnetic field strength will be weaker since the electrons are mirrored back before they hit the dense chromosphere. In the present case, as shown in Figure \ref{Fig:position_final}, the magnetic field strength at the southern ribbon is considerably stronger, which could result in the observed X-ray asymmetry. However, in this same scenario, gyrosynchrotron sources would appear brighter at the footpoint with high magnetic field strength. In the present observations, one would therefore expect brighter radio sources near the southern ribbon, in contradiction with the observed radio sources that are concentrated near the northern ribbon. The observation of both, high energy X-ray and radio sources from the flaring northern ribbon with relatively weaker magnetic field strength, hence cannot be explained with magnetic field strength asymmetry of a single loop. However, since the high energy X-ray and radio sources are not co-spatial, and the magnetic topology is complex, these observations further support the interpretation of two entirely different loops with very different magnetic field asymmetries. 

\section{Conclusion}\label{s:conclusion}

We present a detailed multi-wavelength imaging spectroscopy analysis of multiple radio bursts observed by the VLA at 1.65--2.03 GHz during a microflare. We observe six radio bursts simultaneously with impulsive high-energy X-ray emission at 10$-$18~keV. However, microwave bursts and high-energy X-ray sources are not co-spatial. The bursts are spatially and temporally resolved and interpreted as multiple episodes of electron acceleration. The microflare displayed two EUV flare ribbons, located near opposite magnetic polarities.  All six microwave burst sources and almost all high-energy X-ray sources are asymmetrically located near northern ribbon. The strong asymmetry of the observed radio and X-ray sources is likely due to the complex magnetic topology and multiple loops. The inferred magnetic field strengths during bursts A to C suggest a low coronal origin of the emission. 

Further, the nonthermal electron distributions obtained by fitting the X-ray spectrum and gyrosynchroton spectrum were distinctly different with the low energy cut-off and nonthermal electron density having the biggest discrepancy. 
The nonthermal power obtained for the radio bursts is higher than the one from the X-ray fits by one order of magnitude. However, both values lie within the range of nonthermal powers 
found in a statistical analysis of sub-C class flares by \citet{2008ApJ...677..704H}.

Overall, the contrasting spectral properties and spatial displacements suggest two distinct electron populations.
 This microflare study demonstrates that even microflares can exhibit complex characteristics and behaviours and underlines  the importance and necessity of simultaneous radio and X-ray imaging-spectroscopy observations for our understanding of solar flares, as in the present combination of VLA with RHESSI or, in the future, instruments such as VLA, EOVSA, and STIX on Solar Orbiter. 
\acknowledgments
This paper makes use of public archival data from the VLA observing program VLA/11B-129. The authors acknowledge Tim Bastian, Richard Perley, Michael Rupen, Ken Sowinski, Stephen White for their help in carrying out the observing program. The National Radio Astronomy Observatory is a facility of the National Science Foundation operated under cooperative agreement by Associated Universities, Inc. RS acknowledges support from the Swiss National Science foundation (grant no. 200021\_175832). B.C. and S.Y. are supported by NSF grant AGS-1654382 and NASA grants NNX17AB82G, 80NSSC20K1318 to NJIT. We thank Gregory Fleishman for help with magnetic extrapolation in gx\_simulator software.

\appendix

\section{Optimization of gyrosynchrotron spectral fitting}
Fitting a gyrosynchrotron spectrum is a multi-variable optimization problem making computationally fast fitting of an observed spectrum non-trivial. However, various optimization algorithms offer a wide range of choices to build confidence in the fitted parameters. For the present radio bursts, we investigate the fit results obtained from various algorithms using different techniques. We compare optimization via the MCMC (discussed in the main text), Basinhopping, Brute-force, Differential Evolution and Least-Squares algorithms. 
Among these, only MCMC provides uncertainties on the optimised parameters, i.e. it was preferred for interpretation purpose over the other algorithms. However other algorithms also have unique ways to converge on the global minimum, and offer other advantages. A brief description of each algorithm is given below.

\begin{enumerate}
    \item Basinhopping \citep{lmfit} is a two step procedure. In the first step, the fit function is minimised and a local minimum is calculated. In the second step a random jump is given to the parameters pushing them out of local minimum. After a sufficiently large number of iterations the parameters will converge to the global minimum. This optimization technique works well for deep global minima having rugged structures or valley-like features.     
    \item Differential Evolution optimises the parameters by creating many solution spaces using the constraints of the problem. Here, no gradients are computed. Rather the optimization relies on the measure of quality for a solution space. Since this optimization does not involve gradients, it is particularly suitable for discrete or non-continuous data.
    \item Least-Squares optimization is the standard method, which relies on residuals to find the global minimum. It is computationally lean and user-friendly. 
    \item The Brute-force method divides the parameter space into multi-dimensional grids and computes values of the function on them. The function's values on the multi-dimensional grid are used to calculate the global minimum. 
    \item MCMC algorithm optimization uses probability densities of the parameters and relies on Bayesian inference for optimization \citep{2013PASP..125..306F}. 
\end{enumerate}

\begin{figure*}
\centering
\includegraphics[width=0.6\textwidth]{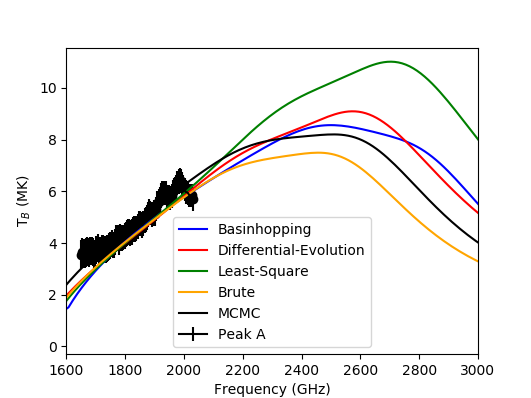}  
\caption{Fit results obtained for radio burst peak A using different algorithms. The `+' plots the datapoints for radio burst A, while the colored lines are the fit for various algorithms.}
\label{Fig:allgsfit}
\end{figure*}

We use the python-based \texttt{lmfit} \citep{lmfit} package to implement all the optimization algorithms.
Table \ref{Tab:gs_methods} lists the fitted parameters for different algorithms. Figure \ref{Fig:allgsfit} shows the brightness temperature fits for all the tested algorithms. The frequency of the peak of the spectrum determines the magnetic field and it increases at higher frequencies. Since the current event is weak and limited in frequency coverage, the fitted magnetic field varies between different algorithms. The minimum and maximum values are 150 and 172 G for Brute-force and Differential-Evolution algorithms respectively, while the MCMC values lie in between. The parameters for Basinhopping and Brute-force are larger than MCMC results. Since Basinhopping depends on abrupt jumps to converge, it can be far from the true global minimum especially when the global minimum is shallow (Figure \ref{Fig:allgsfit}). The Brute-force algorithm is ideal as it samples the entire parameter space. However, since the optimization problem is non-linear, a linear sampling can miss the global minimum, especially when the global minimum is relatively narrower than the sampling in the parameter space. In this approach, the sampling must be high, along with a sufficiently large range of the parameter values. However, these constraints make the Brute-force method computationally slow for an average computer user. The Differential-Evolution and Least-Squares methods gave mostly consistent results with MCMC. However, nonthermal densities from MCMC are slightly lower than the estimates from other methods. Like the magnetic field, increasing nonthermal density shifts the spectrum towards higher frequency. Therefore, uncertainty in determining the peak will impact the fits estimates. In terms of computation times, MCMC and Brute-force were slower than the other algorithms on a single processor. However, overall the fits from all the algorithms did not give any unreasonable estimates of the physical parameters. All of them are quite consistent with each other, within a factor of few.
\begin{table*}[]
    \centering
    \begin{tabular}{|c|c|c|c|c|}
    \hline
    Method & B (Gauss) & n$_{b}$ ($\times$10$^8$ cm$^{-3}$) &  E$_{low}$ (keV) & $\delta$  \\
    \hline
    Basinhopping & 155 & 11.7 & 3.4 & 6.3 \\
    Brute-force & 150 & 11.0 & 4.0 & 6.6 \\
    Differential-Evolution & 172 & 6.2 & 2.7 & 5.6 \\
    Least-Squares & 160 & 2.2 & 4.0 & 6.5\\
    MCMC & 158.6$\pm$12.6& 8.8$\pm$6.4 & 4.1$\pm$0.9 & 5.9$\pm$0.6\\
    \hline
    \end{tabular}
    \caption{Fitted parameters from different algorithms for radio burst A.}
    \label{Tab:gs_methods}
\end{table*}

\bibliography{manuscript.bib}



\end{document}